\documentclass[12pt,reqno]{amsart}
\RequirePackage{multirow}
\usepackage{xifthen,setspace,multirow}
\usepackage{diagbox,graphicx}
\usepackage{caption,subcaption}
\usepackage{float}
\usepackage[margin=3cm]{geometry}
\usepackage{bbm}
\usepackage{xcolor}
\usepackage{enumitem}
\usepackage{bm}
\usepackage[linesnumbered,ruled,vlined]{algorithm2e}
\usepackage{hyperref}
\usepackage{esvect}
\usepackage{amsmath,amssymb,mathrsfs} 
\usepackage{lipsum}  
\newtheorem{thm}{Theorem}

\newtheorem{definition}{Definition}
\newtheorem{lem}{Lemma}

\newcommand{\mx}{{\mathbf{X}}}

\renewcommand{\leq}{\leqslant}

\newcommand{\e}{{\mathbb{E}}}

\newcommand{\R}{{\mathbb{R}}}

\newcommand{\p}{{\mathbb{P}}}

\allowdisplaybreaks

\title[Graph-Based Feature Selection]{Feature Selection in  High-dimensional Spaces Using Graph-Based Methods}

\author[Ghosh]{Swarnadip Ghosh}
\thanks{The first two authors contributed equally to the paper.}
\address{email {\tt gswarnadip@gmail.com}}

\author[Mukherjee]{Somabha Mukherjee}
\address{email {\tt somabha@nus.edu.sg}}

\author[Agarwal]{Divyansh Agarwal} 
\address{email {\tt dagarwal@mgh.harvard.edu}}

\author[He]{Yichen He}
\address{email {\tt e0732876@u.nus.edu}}

\author[Song]{Mingzhi Song}
\address{email {\tt e0732845@u.nus.edu}}

\author[Pei]{Xuejiao Pei}
\address{email {\tt e0729791@u.nus.edu}}

\begin{document}

	\begin{abstract} 
		High-dimensional feature selection is a central problem in a variety of application domains such as machine learning, image analysis, and genomics. In this paper, we propose graph-based tests as a useful basis for feature selection. We describe an algorithm for selecting informative features in high-dimensional data, where each observation comes from one of $K$ different distributions. Our algorithm can be applied in a completely nonparametric setup without any distributional assumptions on the data, and it aims at outputting those features in the data, that contribute the most to the overall distributional variation. At the heart of our method is the recursive application of distribution-free graph-based tests on subsets of the feature set, located at different depths of a hierarchical clustering tree constructed from the data. Our algorithm recovers all truly contributing features with high probability, while ensuring optimal control on false-discovery. We show the superior performance of our method over other existing ones through synthetic data, and demonstrate the utility of this method on several real-life datasets from the domains of climate change and biology, wherein our algorithm is not only able to detect known features expected to be associated with the underlying process, but also discovers novel targets that can be subsequently studied.
	\end{abstract}



	\maketitle
	
	\section{Introduction}\label{int}
	
	High dimensional statistical problems arise frequently across diverse fields of scientific research. Variable selection plays a pivotal role in such settings.
	For example, with an estimated 30,000 genes in the human genome, computational investigations in genomics often seek to identify the features of utmost importance, either to a disease, developmental or physiological process. In such high-dimensional data analyses, robust and scalable methods for feature selection are of paramount importance \cite{bioinform}. 
	Several methods have been developed for high dimensional linear regression such as the lasso(see \cite{tib96}).
	The quality of a feature selection mechanism is often compromised under model misspecifications by the high-dimensional nature of the data. When there are a large number of nuisance parameters in a misspecified model, parametric methods such as regression often perform poorly, especially so in high-dimensional setups.  This necessitates the development of non-parametric methods, which make no distributional assumptions on the data, but are still powerful for a wide class of alternatives.
	
	In this paper, we are interested in the problem of variable selection in the following distributional context. Suppose we have data coming from several multivariate distributions, and we want to detect any non homogeneity across these distributions. In Statistics, this scenario is addressed by the classical $K$ sample problem where one concludes that $K$ multivariate distributions are non-identical based on some test statistics. Our target is to understand which marginals of these multivariate distributions contribute the most to their overall change, if any.
	
	Understanding the important variables for a $K$ sample problem that cause a distributional difference is of practical importance as it can provide actionable features for further investigation. For univariate data,  there are several celebrated distribution-free two-sample tests such as the Kolmogorov-Smirnov maximum deviation test \cite{smirnov}, the Wald-Wolfowitz runs test \cite{ww}, and the Mann-Whitney rank-sum test \cite{mann_whitney} (see \cite{gc} for more on these tests). In the multivariate scenario, distributional variations in the dataset are often amplified by the joint effects of some closely related features, rather than individual effects examined in isolation. The multivariate dependency structure motivated researchers to generalize univariate methods to higher dimensions (See \cite{weiss,bickel}). 
	Friedman and Rafsky \cite{fr} proposed a multivariate two-sample test (multivariate run test) to test homogeneity based on the minimal spanning tree\footnote{Given a finite set $S \subset \R^d$, the {\it minimum spanning tree} (MST) of $S$ is a tree  with vertex-set $S$, which has the minimum value of sum of the Euclidean lengths of its edges.} of the sample points, and Schilling \cite{schi} proposed a  multivariate two-sample test based on the nearest neighbor graph. Even though many of these methods can be effectively used in high-dimensional problems, none of them inherit the exact distribution-free property of the univariate tests. Biswas et al. \cite{biswas} proposed another two-sample test based on Hamiltonian cycles, which is distribution-free in finite samples. However, computing the minimum weight Hamiltonian path is NP-hard, making this test computationally prohibitive beyond small sample sizes. 
	
	In this work we use the exact distribution free multisample test developed in \cite{mukherjee}, a generalization of the multivariate two-sample {\it crossmatch test} introduced by Rosenbaum \cite{rosen}, to test homogeneity across multiple samples. This test is based on the number of between-sample edges in the minimum non-bipartite matching graph constructed on the pooled sample. This method is computationally efficient since the test statistic can be computed in time which is polynomial in both the number of samples and the dimension of the data. We perform the test in a hierarchical way with a stopping rule based on adjusted $p$ value to identify the important set of variables in this $K$ sample framework. For this purpose, we use the hierarchical clustering approach proposed in \cite{hierarchical}. This method begins by testing distributional homogeneity across all the variables. Depending on whether there is enough evidence of non-homogeneity, the method then  proceeds by testing homogeneity across smaller subsets of variables and may continue till individual variables are reached. The smallest possible sets of variables, which still exhibit a significant evidence of distributional difference, are retained.
	The final outcome of this method is a subset of variables, which are minimally significant, i.e., do not contain redundant non-significant variables. Since this method is based on graph-based multisample tests, we call it the \textit{graph-based feature selection} (GFS) algorithm. 
	With increasing sample size, the proportion of true signals recovered by the GFS algorithm approaches $1$, while at the same time ensuring a control on the probability of discovering at least one false signal.
	

	
	The rest of the paper is organized as follows. We begin by describing the GFS algorithm in Section~\ref{sec:meth}. Section ~\ref{theory} gives theoretical guarantees of our algorithm. In section ~\ref{sec:sim}, we provide numerical experiments under a variety of setups with varying location and scale parameters, and demonstrate the efficacy of the GFS in such cases. The procedure is then applied on a forest fires data, a single cell transcriptomics data, a mice protein data, and a brain cancer data in Section \ref{realdata}. Section ~\ref{sec:discussion} concludes the paper with a discussion. Finally, the Appendix provides proofs of the lemmas quoted in the main body of the manuscript. 
	
	
	\section{Methodology}\label{sec:meth}
	In this section, we describe our graph-based feature selection (GFS) algorithm. Suppose that we have $K$ unknown multivariate probability distributions $F_1,\ldots,F_K$ on $\mathbb{R}^d$. For each $1\leq i \leq K$, suppose that we have access to $n_i$ observations from the distribution $F_i$. We assume that all the $n:=n_1+\ldots+n_K$ observations are independent. Our goal is to select those marginals of $F_1,\ldots,F_K$, which contribute the most to the difference in these distributions, if there is any.
	
	The $n$ observations are stacked along the rows of an $n\times d$ data matrix $\mathbf{X}$, with the first $n_1$ rows of $\mathbf{X}$ denoting the observations coming from $F_1$, the next $n_2$ rows of $\mathbf{X}$ denoting the observations coming from $F_2$, and so on. For a set $S \subseteq [d]:= \{1,\ldots,d\}$, let $\mx_S$ denote the submatrix of $\mx$ formed by all the rows of $\mx$ and only those columns of $\mx$ whose indices are in $S$. Hence, $\mx_S$ is an $n\times |S|$ matrix. Suppose that we have a non-parameteric test $T$ that can test 
	$$H_0: F_1=\ldots =F_K\quad\quad\textrm{v.s.}\quad\quad H_1: F_i \neq F_j~\textrm{for some}~1\leq i<j\leq K$$
	based on a data matrix of observations from $F_1,\ldots F_K$. A natural choice of $T$
	is any non-parametric distribution-free graph-based test for equality of multivariate distributions (for example, the MCM and the MMCM tests in \cite{mukherjee}). 
	
	The algorithm starts by constructing a tree $\mathcal{T}$ of clusters, satisfying the following properties:
	\begin{enumerate}
		\item The root node of $\mathcal{T}$ is the set $[d]$.\label{en1}
		\item Every non-leaf node $C$ of $\mathcal{T}$ has two children, which are disjoint sets uniting to $C$. \label{en2}
		\item The leaf nodes of $\mathcal{T}$ are singleton sets.\label{en3}
	\end{enumerate} 
	The hierarchical structure of $\mathcal{T}$ implies that for any two nodes $C$ and $D$ of $\mathcal{T}$, either $C \subseteq D$ or $D \subseteq C$ or $C\bigcap D = \emptyset$.
	
	A natural way to construct such a tree $\mathcal{T}$ is to use a hierarchical clustering procedure like Single Linkage agglomerative clustering (see \cite{hastie}, \cite{hcscheme}, \cite{hcjspi}). Some other hierarchical clustering methods can be found in \cite{treeform}, \cite{morteza}. The starting point of this procedure is a matrix $D \in \mathbb{R}^{d\times d}$ of pairwise dissimilarities between the variables $1,\ldots,d$. In the simulation and real data analysis parts of this paper, we take $D = J-R$, where $R$ is the correlation matrix of the $d$ variables and $J$ is the matrix of all ones. The tree is built in a sequential way. In the beginning of the agglomerative clustering process, each element is in a cluster of its own. The clusters are then sequentially combined into larger clusters, until all elements end up being in the same cluster. At each step, the two clusters separated by the shortest distance are combined. For Single Linkage agglomerative clustering, the distance between two clusters is defined as the minimum distance between elements of the two clusters.
	
	Once the tree $\mathcal{T}$ is constructed, the next step is to get hold of some \textit{minimally significant} subsets of features. Towards this, we need a couple of definitions.
	
	\begin{definition}\label{adjp}
		The adjusted $p$-value of a node $C$ in the tree $\mathcal{T}$ is defined as:
		$$p_{\mathrm{adj}}^C := p^C \frac{d}{|C|}~,$$ where $|C|$ denotes the cardinality of $C$, and $p^C$ denotes the $p$-value of the test $T$ conducted on the matrix $\bm X_C$.
	\end{definition}
	
	\begin{definition}\label{terminal}
		For $\alpha \in [0,1]$, a node $C$ in the tree $\mathcal{T}$ will be called an $\alpha$-terminal node, if the following three conditions are satisfied
		
		\begin{enumerate}
			\item $p_{\mathrm{adj}}^C \le \alpha$;
			
			\item $p_{\mathrm{adj}}^B \le \alpha$ for every ancestor $B$ of $C$ in the tree $\mathcal{T}$;
			
			\item $p_{\mathrm{adj}}^B > \alpha$ for every child $B$ of $C$ in the tree $\mathcal{T}$.
		\end{enumerate}
	\end{definition}
	Note that condition (2) in Definition \ref{terminal} is satisfied vacuously by the root node of $\mathcal{T}$, and condition (3) is satisfied vacuously by the leaf nodes of $\mathcal{T}$. Our final set of selected features at level $\alpha$, is the union of all the $\alpha$-terminal nodes of $\mathcal{T}$. Below, we give an algorithmic implementation of our feature selection procedure.
	
	\bigskip
	\begin{algorithm}[H]\label{alg1} 
		\KwIn{A tree $\mathcal{T}$ satisfying (\ref{en1}), (\ref{en2}) and (\ref{en3})}
		\KwOut{A set $S \subseteq [d]$ of selected variables}
		\textbf{Initialization:} Set $C$ to be the root node of $\mathcal{T}$ and $S = \emptyset$\;
		Perform the test $T$ on $\bm X_C$ and report the adjusted $p$-value $p_{\mathrm{adj}}^C$. If $p_{\mathrm{adj}}^C > \alpha$ and $C = [d]:=\{1,2,\ldots,d\}$, stop the algorithm; else if $p_{\mathrm{adj}}^C \le \alpha$ and $|C|=1$, set $S \leftarrow S\bigcup C$; else if $p_{\mathrm{adj}}^C \le \alpha$ and $\min\{p_{\mathrm{adj}}^{C_1}, p_{\mathrm{adj}}^{C_2}\} > \alpha$ for children $C_1, C_2$ of $C$, set $S \leftarrow S \bigcup C$; else go to Step 3\;
		Perform Step 2 with $C$ replaced by every child $B$ of $C$ satisfying $p_{\mathrm{adj}}^B \le \alpha$.

		\caption{Feature Selection Algorithm}
	\end{algorithm}
	\bigskip
	Algorithm \ref{alg1} is implemented as a binary recursion in the \textsf{R} package \textsf{GFS} (under preparation).
	
	\section{Theoretical Guarantees of the GFS Algorithm}\label{theory}
	In this section, we prove some theoretical guarantees of the GFS algorithm. In order to do so, we need to define a precise notional framework for features/marginals of multivariate distributions that contribute to their difference. To start with, the null hypothesis for a set of features $L \subseteq [d]$ is defined as:
	$$H_{0,L}: F_{1[L]}=\ldots =F_{K[L]}$$ 
	where $F_{i[L]}$ denotes the joint distribution of $(X_j)_{j\in L}$ where $X:=(X_1,\ldots,X_d) \sim F_i$. We call a set of features $L \subseteq [d]$ a \textit{signal set}, if:
	\begin{enumerate}
		\item $H_{0,L^{c}}$ is true. \label{cc1}
		\item $\forall~ L^{\prime}\subseteq L$, $H_{0,L^{\prime}}$ is not true.\label{cc2}
	\end{enumerate}
	
	It follows from Lemma \ref{unique1} that a signal set is unique. We can now define the familywise error rate (FWER), false discovery rate (FDR) and power of a feature selection procedure.
	
	\begin{definition}\label{fwfdp}
		Let $L$ be the signal set. The FWER, FDR and power of a feature selection procedure reporting a (random) set of selected features $S$ are defined as:
		\begin{enumerate}
			\item $\mathrm{FWER} := \p(S\setminus L \ne \emptyset);$
			\item $\mathrm{FDR} := \e \left(|S\setminus L|/\max\{|S|,1\}\right);$
			\item $\mathrm{Power} := \p(L \subseteq S).$ 
		\end{enumerate}
	\end{definition}
	
	\begin{thm}\label{mainthm}
		Suppose that $d$ is fixed as $n \rightarrow \infty$. Then, in the asymptotic regime where the underlying test $T$ is consistent at every level, the power of the GFS algorithm goes to $1$, whereas the FWER and FDR is controlled at level $\alpha + o(1)$.
	\end{thm}
	
	\begin{proof}
		Let $L$ be the signal set and $S$ be the set of selected features. We start by showing that the power of the GFS algorithm goes to $1$. Towards this, note that for every set $B$ containing some $\ell \in L$, $H_{0,B}$ is not true, and hence, by the consistency of the test $T$ at every level, we have:
		$$\p(p_{\mathrm{adj}}^B \le \beta) \rightarrow 1\quad\textrm{for all}~\beta > 0$$
		which is same as saying that $p_{\mathrm{adj}}^B \rightarrow 0$ in probability.
		Since $d$ is fixed, we have:
		\begin{equation}\label{maxpower}
			\max \{p_{\mathrm{adj}}^B: \ell \in B~\textrm{for some}~ \ell \in L\} \rightarrow 0\quad\textrm{in probability}~.
		\end{equation}
		It follows from \eqref{maxpower} that 
		$$\max \{p_{\mathrm{adj}}^B:~B = \{\ell\}~\textrm{or}~B~\textrm{is an ancestor of}~\{\ell\}~\textrm{in}~\mathcal{T}~\textrm{for some}~\ell \in L\}\rightarrow 0\quad\textrm{in probability}.$$
		This shows that 
		\begin{equation}\label{f1}
			\p(\textrm{every}~\{\ell\} \subseteq L~\textrm{is an}~\alpha\textrm{-terminal node}) \rightarrow 1~.
		\end{equation}
		Since $S$ was defined as the union of all $\alpha$-terminal nodes, it follows from \eqref{f1} that $\p(L \subseteq S) \rightarrow 1$, i.e. the power of the GFS algorithm goes to $1$.
		
		Next, we show that the FWER of the GFS algorithm is $\alpha + o(1)$. Towards this, suppose that $S \setminus L \ne \emptyset$, and choose $v \in S \setminus L$. Since $v \in S$, $v \in V$ for some $\alpha$-terminal node $V$ of $\mathcal{T}$. Now, there may be two cases. If $V \subseteq L^c$, then by (\ref{cc1}), $H_{0,V}$ is true. Also, since $V$ is an $\alpha$-terminal node, $$\max_{U \in \mathcal{T}: U \supseteq V} p_{\mathrm{adj}}^{U} \le \alpha.$$ Hence, $V \in \mathcal{T}_0 \bigcap \mathcal{T}_{\mathrm{rej}}$, where $\mathcal{T}_0$ and $\mathcal{T}_{\mathrm{rej}}$ are defined in the statement of Lemma \ref{ap2}. This implies that $\mathcal{T}_0 \bigcap \mathcal{T}_{\mathrm{rej}} \ne \emptyset$. Otherwise, $V$ contains some $\ell \in L$. Thus, $|V| \ge 2$, and hence,  $V$ has two non-empty children $V_1$ and $V_2$. By (\ref{en2}), $\ell \in V_i$ for some $i \in \{1,2\}$. By (\ref{cc2}) $H_{0,\{\ell\}}$ is not true, and hence, $H_{0,V_i}$ is not true. Also, since $V$ is an $\alpha$-terminal node, $p_{\mathrm{adj}}^{V_i} > \alpha$. Hence, we have:
		$$\{S \setminus L \ne \emptyset\} \subseteq \{\mathcal{T}_0 \cap \mathcal{T}_{\mathrm{rej}} \ne \emptyset\}~ \bigcup \underset{\substack{
				V \subseteq [d]: H_{0,V}\\ ~\textrm{is not true}}}{\bigcup} \{p_{\mathrm{adj}}^{V} > \alpha\}~.$$
		By Lemma \ref{ap2}, $\p(\mathcal{T}_0 \cap \mathcal{T}_{\mathrm{rej}} \ne \emptyset) \le \alpha$. Also, by consistency of $T$, we know that $\p(p_{\mathrm{adj}}^{V} > \alpha) = o(1)$ for all $V \subseteq [d]$ such that $H_{0,V}$ is not true. Since $d$ is fixed, it follows that $\p(S \setminus L \ne \emptyset) \le \alpha + o(1)$. The FDR control at level $\alpha$ follows from the fact that $FDR \leq FWER.$ This completes the proof of Theorem \ref{mainthm}. 
		
	\end{proof}
	
	\section{Simulations}\label{sec:sim}
	
	In this section, we show the performance of our feature selection algorithm 
	when the test $T$ is the MMCM test (see \cite{mukherjee}), and the underlying distributions are assumed to be taken from a location and a scale family. The MMCM test starts by constructing a vector $\bm A := ((A_{ij}))_{1\le i<j\le K}$ of cross-counts, where $A_{ij}$ denotes the number of edges in the minimal non-bipartite matching graph\footnote{The minimal non-bipartite matching graph on $n$ nodes is defined as a matching on these $n$ points that has minimum value of the sum of Euclidean lengths of edges.} constructed on the $n$ sample points, that have one endpoint coming from $F_i$ and the other endpoint from $F_j$. It was shown in \cite{mukherjee} that the vector $\bm A$ is distribution free under the null hypothesis of distributional equality. Using this fact, the MMCM test statistic is defined as:
	$$T := (\bm A - \e_{H_0} \bm A)^\top \mathrm{Cov}_{H_0}^{-1} (\bm A) (\bm A - \e_{H_0} \bm A)$$
	where $\e_{H_0}$ and $\mathrm{Cov}_{H_0}$ denote the expectation and covariance operators under the null hypothesis of distributional equality. The null hypothesis is rejected for large values of $T$. Although this test is distribution-free under the null, its asymptotic $\chi_{\binom{K}{2}}^2$ distribution is often more useful in practice for choosing the exact rejection threshold.
	
	Each of the location and scale settings is executed in the distributional framework when the underlying data is generated from a multivariate normal distribution. A typical case looks as follows. The $d$-variate normal distributions are perturbed in randomly chosen $s$ marginals $j_1,j_2,\cdots,j_s$, and the remaining $d-s$ marginals are taken to be standard Gaussians. In the location setting, the perturbations in the $s$ marginals are in the means, with variances fixed at $1$, while in the scale setting, the perturbations in the $s$ marginals are in the variances, with means fixed at $0$. In each setting, we run $R$ replications of this experiment, resulting in $R$-many tuples of selected variables. We then report the empirical versions of the FWER, FDR and power of the GFS algorithm conducted at $5\%$ level, over the $R$ replications.
	\subsection{The Location Setting:}\label{loc1}
	In this setting, we start with a fixed vector $\boldsymbol{\mu}$ of length $d$. We then randomly choose $s$-many entries of $\boldsymbol{\mu}$ and make them all equal to some $\theta$. The remaining entries of $\boldsymbol{\mu}$ are set at $0$. The distribution $F_i$ is then taken to be the multivariate Gaussian with mean $i\boldsymbol{\mu}$ and identity covariance matrix for $1\leq i\leq K$. We vary $\theta$ in the set $\{0.15,0.2,\cdots,0.70,0.75\}$ and take $d = 100$, $s=25$, $R=100$. We set the number of classes $K=5$, with all class sizes $= 200$, and report the empirical FWER, FDR and power over $R$ replications. The graphs of the empirical FWER, FDR and power are given below, and the exact numerics are presented in Appendix \ref{num11}.
	
	\subsection{The Scale Setting:}\label{sc1}
	In this setting, we start with a fixed positive real number $\theta$. We then randomly choose $s$-many indices from $\{1,\ldots,d\}$, and define $F_i$ as a $d$-variate normal distribution with mean zero, and covariance matrix obtained by replacing the $s$ many chosen diagonal entries of the identity matrix, by $1+(i-1)\theta$. We vary $\theta$ in the set $\{1,1.5,2,2.5\cdots,14,14.5,15\}$ and take $d=100$, $s=25$, $R=100$. We set the number of classes $K=5$, with class sizes $=200$.

	
	\begin{figure}
		\begin{center}
			\includegraphics[height=4in,width=4.6in]{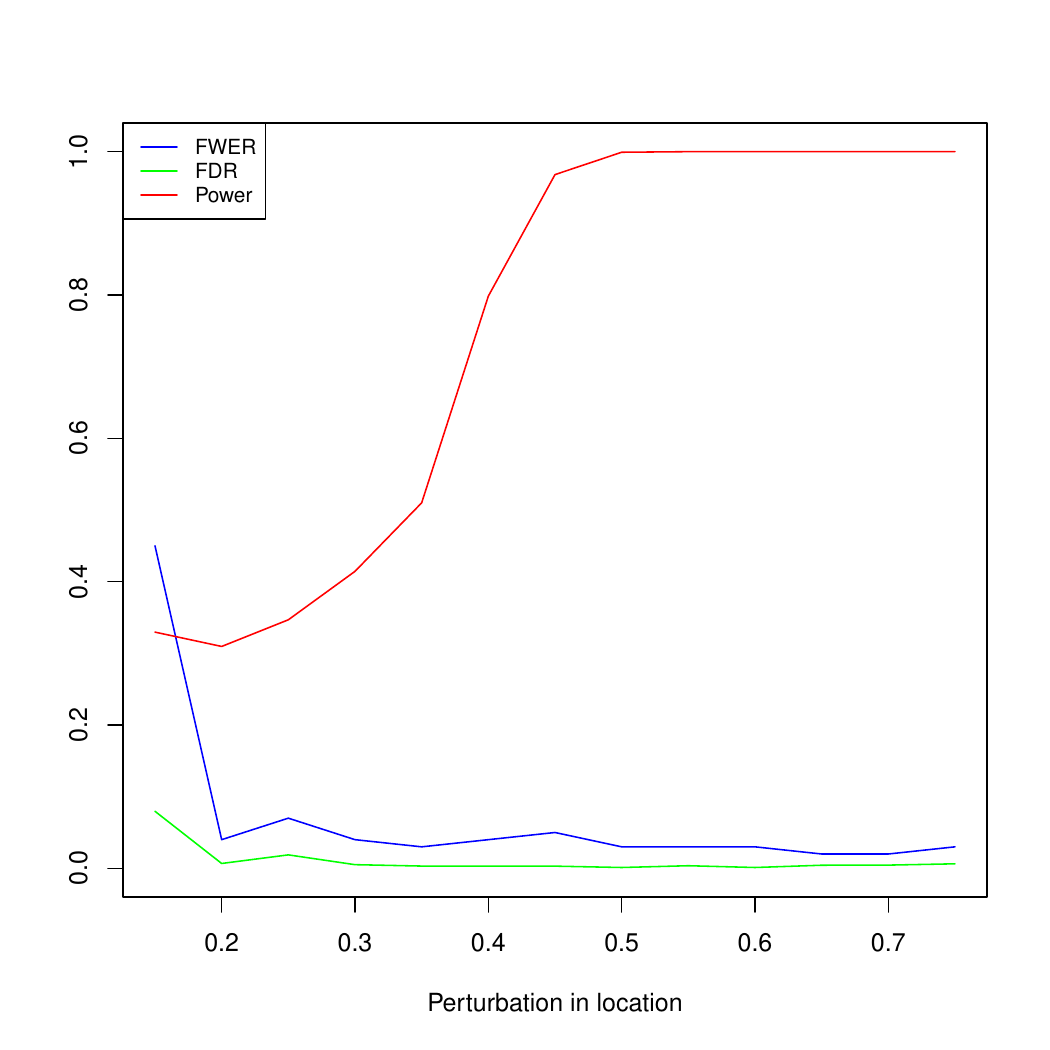}
		\end{center}
		\caption{FWER, FDR and power of the GFS algorithm using the MMCM test in the location setting
			\label{fig:illustratet1}}
	\end{figure}	
	
	\begin{figure}
		\begin{center}
			\includegraphics[height=4in,width=4.6in]{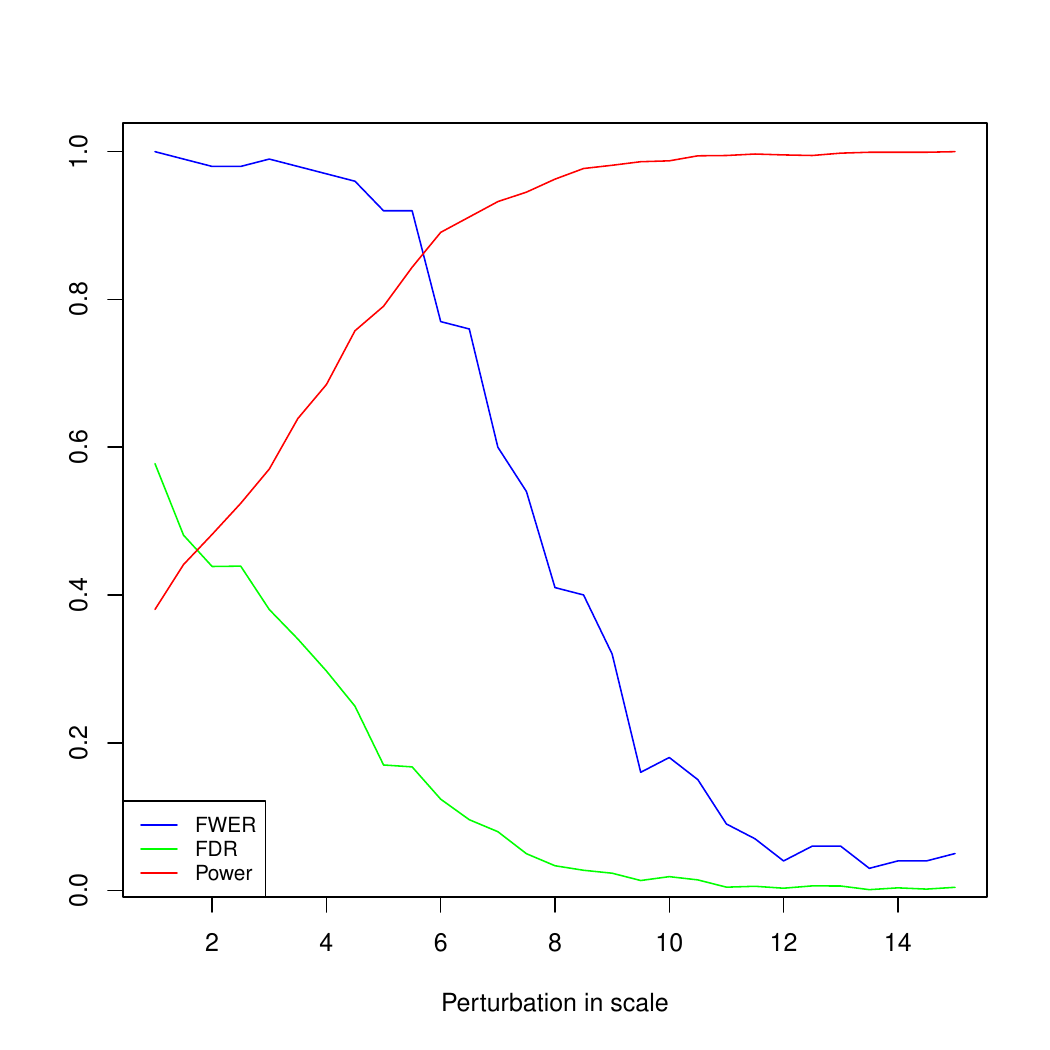}
		\end{center}
		\caption{FWER, FDR and power of the GFS algorithm using the MMCM test in the scale setting
			\label{fig:illustratet2}}
	\end{figure}
	
	Figures \ref{fig:illustratet1} and \ref{fig:illustratet2} show that the (empirical) power of the GFS algorithm increases to $1$ as the perturbation in location/scale increases, whereas the FDR and FWER decrease sharply, and fall below the $5\%$ level with increase in perturbation. 

	\subsection{Comparison with the Kruskal Wallis test in location-scale family with dependence:}\label{kw1}
	
	For this simulation we consider a similar set up as before. We take $F_i$ to be a $d$-variate normal distribution with mean $\mu^{(i)}$ and covariance matrix $\Sigma^{(i)}$ of the following form:
	

	$$\mu^{(i)} = \begin{cases} 0 \mbox{ if } i \not \in L\\
		i\theta \mbox{ if } i \in L\\
	\end{cases} \quad
	\Sigma_{jk}^{(i)} = \begin{cases}
		\rho,~ \mbox{ if }  j \not \in L \mbox{ and } k \not \in L \mbox{ and } j \neq k\\
		1,~ \mbox{ if }  j \not \in L \mbox{ and } k \not \in L \mbox{ and } j = k\\
		\rho (1+ (i-1)\phi) ~\mbox{ if }  j \in L \mbox{ and } k \in L \mbox{ and } j \neq k\\
		1+ (i-1)\phi, ~\mbox{ if }  j \in L \mbox{ and } k \in L \mbox{ and } j = k\\
		\rho,~ \mbox{otherwise}\\
	\end{cases}
	$$
	where $L$ is a randomly chosen subset of $\{1,\ldots,d\}$. We take $\theta \in \lbrace 0, 0.1, 0.2, \cdots, 0.7\rbrace$ and $\phi \in \lbrace 0, 1, 3, 5, \cdots, 15 \rbrace$ for $\rho = 0.1$. $\theta = 0$ corresponds to no shift in location and $\phi = 0$ corresponds to no shift in scale.
	We compare our method with a test developed by Kruskal and Wallis \cite{ks,kswallis}, a distribution-free test, which is a K sample analogue of the Mann-Whitney test. To select the set of important variables using the Kruskal Wallis test, we performed a univariate test and did a multiplicity correction. We plot the empirical FWER, FDR and power calculated over $30$ replicates.

	\begin{figure}
		\centering
		\begin{subfigure}{.4\textwidth}
			\centering	
			\includegraphics[width=1\linewidth]{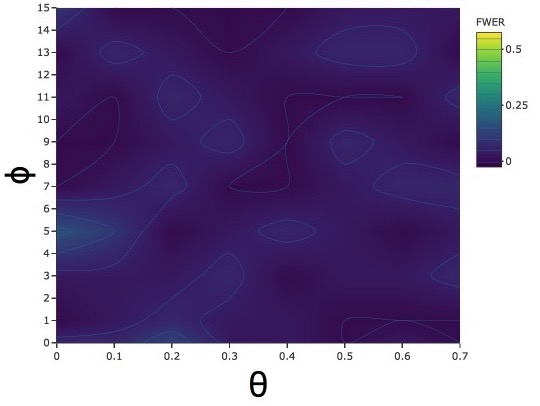}
		\end{subfigure}
		\centering
		\begin{subfigure}{.4\textwidth}
			\centering	
			\includegraphics[width=1\linewidth]{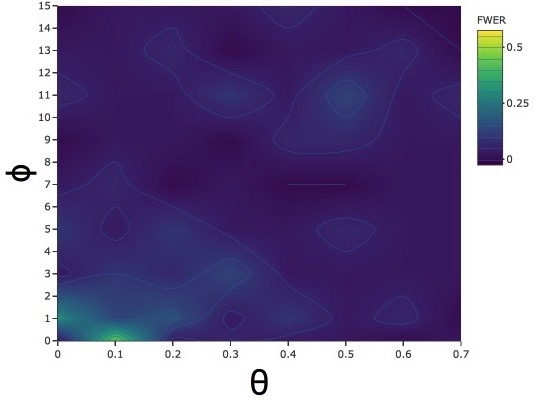}
		\end{subfigure}
		\caption{Contour plot of FWER of Kruskal Wallis (left panel) and the GFS algorithm (right panel)}
		
	\end{figure}

	\begin{figure}
		\centering
		\begin{subfigure}{.4\textwidth}
			\centering	
			\includegraphics[width=1\linewidth]{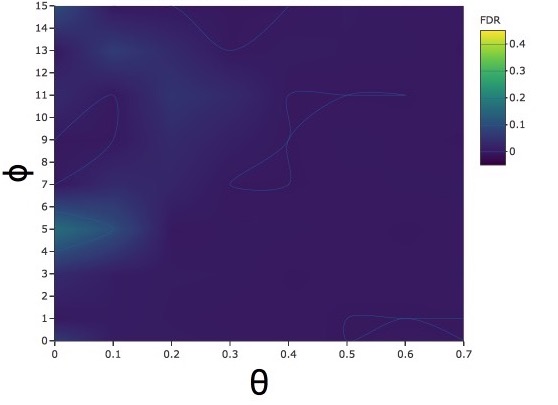}
		\end{subfigure}
		\centering
		\begin{subfigure}{.4\textwidth}
			\centering	
			\includegraphics[width=1\linewidth]{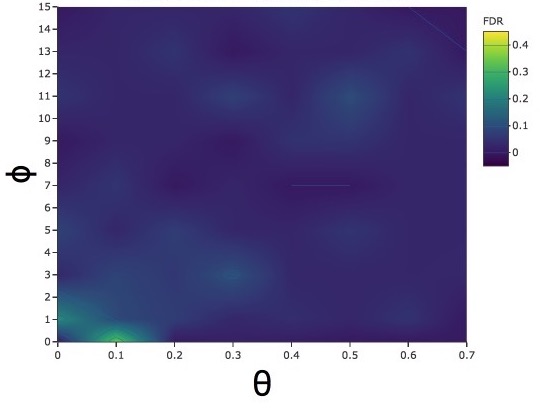}
		\end{subfigure}
		\caption{Contour plot of FDR of Kruskal Wallis (left panel) and the GFS algorithm (right panel)}
		
	\end{figure}
	
	\begin{figure}
		\centering
		\begin{subfigure}{.4\textwidth}
			\centering	
			\includegraphics[width=1\linewidth]{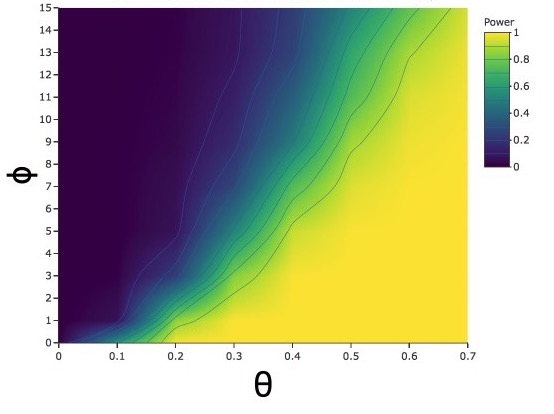}
		\end{subfigure}
		\centering
		\begin{subfigure}{.4\textwidth}
			\centering	
			\includegraphics[width=1\linewidth]{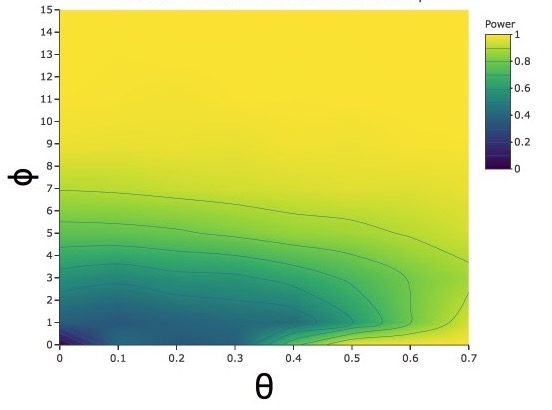}
		\end{subfigure}
		\caption{\label{fig:power}Contour plot of power of Kruskal Wallis (left panel) and the GFS algorithm (right panel)}
		
	\end{figure}
	
	We observe that the Kruskal Wallis test does not detect any change in scale. However, in absence of any scale change, it performs better in detecting a location shift, even in low signal domain. This is expected as the Kruskal Wallis test is powerful in location shift set ups only.

	Next we consider a simulation set up where the samples do not come from the same parametric family.
	
	\subsection{Comparison with the Kruskal Wallis test in different parametric families:}
	
	In this set up we take $K = 3$, i.e., we assume that the data is coming from $3$ different classes. 
	For each of the distributions $F_1, F_2$ and $F_3$, the last $d - \lceil \frac{d}{4} \rceil$ marginals are taken to be independent standard Gaussian. As for the first $\lceil \frac{d}{4} \rceil$ marginals they are assumed to be sampled independently from a multivariate standard Gaussian distribution for $F_1$, multivariate standard Cauchy distribution for $F_2$ and multivariate standard t distribution for $F_3$.
	For this simulation, we set each of the class sizes at $400$. We varied $d$ within $\lbrace50, 75, 100,\cdots,250 \rbrace $.
	The degrees of freedom of the multivariate t distribution was taken to be 7. In table \ref{table:differentparametric}, we compare the performance of Kruskal Wallis and our proposed GFS algorithm. 
	
	
	\begin{table}[H]
		\centering
		\begin{table}[H]
			\begin{tabular}{|c||c|c|c|}
				\hline
				$p\downarrow$ &  FWER of GFS & Power of GFS & Power of Kruskal Wallis\\  
				\hline
				\hline
				\multirow{1}{*}{50} & 0.0667& 0.6128 & 0.0000 \\
				\hline
				\multirow{1}{*}{75} & 0.0333& 0.5667 & 0.0000 \\
				\hline
				\multirow{1}{*}{100} & 0.1333&0.5814 & 0.0000 \\
				\hline
				\multirow{1}{*}{125} & 0.0000&0.5667 & 0.0000  \\
				\hline
				\multirow{1}{*}{150} &  0.0667&0.5447 & 0.0009\\
				\hline
				\multirow{1}{*}{175} & 0.0333&0.5326 & 0.0008 \\
				\hline
				\multirow{1}{*}{200} &0.0333& 0.5160 & 0.0000 \\
				\hline
				\multirow{1}{*}{225} & 0.0000&0.5211 & 0.0006  \\
				\hline
				\multirow{1}{*}{250} &  0.0333&0.5138 & 0.0005\\
				\hline
			\end{tabular}
		\end{table}
		\vspace{-0.1in}
		
		\caption{Different parametric family: power}
		\label{table:differentparametric}
	\end{table}
	We do not tabulate FWER and FDR of Kruskal Wallis test as they are nearly 0. 
	
	\section{Real Data Analysis}\label{realdata}
	In this section, we show the performance of our GFS algorithm on several real datasets.
	
	\subsection{Portugal Forest Fires Data}\label{portugal}
	The Portugal forest fires data \textcolor{blue}{\url{http://archive.ics.uci.edu/ml/datasets/Forest+Fires}} (also see \cite{portugalpaper}) describes several meteorological features of a number of locations in  Montesinho natural park, from the Tr\'as-os-Montes northeast region of Portugal, and provides a measure of the areas burnt by forest fires in each of these locations. This dataset contains 517 instances, collected over a period of four years (2000-2004). Whenever a forest fire broke out, the following spatio-temporal and meteorological information were collected: \textit{spatial locations ($x$ and $y$ coordinates)}, \textit{date of occurrence}, \textit{Fine Fuel Moisture Code (FFMC)}, \textit{Duff Moisture Code (DMC)},   \textit{Drought Code (DC)}, \textit{Initial Spread Index (ISI)}, \textit{temperature}, \textit{relative humidity (RH)}, \textit{wind speed}, \textit{rain}, and \textit{area burnt}. 
	
	In our analysis, we included all the above attributes except \textit{date of occurrence} and \textit{area burnt} as relevant features, giving us a total of 10 univariate features. The variable \textit{area burnt} was used as a classifier, and since its histogram has a huge peak at $0$, we discretized its range into two classes: $0$ and positive. We then used the GFS algorithm to identify which of the 10 features would be most informative about whether an area is burnt.
	
	The GFS algorithm picked six of the 10 features, namely all the component variables of the Fire Weather Index (FWI) system: FFMC, DMC, DC, ISI, and the two weather attributes: temperature and wind, as significant variables. On the contrary, when the Kruskal Wallis test was applied to each of the 10 features and a multiple hypothesis testing correction was applied, none of the 10 variable met the threshold of 0.05 for statistical significance. This, in turn, indicates a greater detection power of the multisample graph-based tests in the generality. In fact, the smallest p-value reported by the univariate Kruskal-Wallis test is $0.017$, which is larger than the Bonferroni multiple-comparison level of $0.005$. 
	
	\subsection{Single Cell Transcriptomics Data}\label{sec:genomicdata}
	
	Feature selection is especially important in modern biomedical research, where high-throughput and multi-dimensional datasets are becoming increasingly prevalent with recent advances in single cell technologies \cite{slovin}. For instance, to characterize novel cellular states, single cell RNA sequencing (scRNA-seq) is used in exploratory analyses; scRNA-seq data analysis procedures such as cell type clustering or dimension reduction are often driven by select gene expression profiles. Appropriate feature selection in this domain has major implications on how differences or similarities are estimated between pairs of cells, subsequently affecting any downstream methods \cite{bbab34, hicks}. Selection of differentially expressed genes or highly variable features, therefore, holds utmost importance for both computational efficiency and for drawing meaningful scientific inferences from the data \cite{slovin, bbab34}.
	
	Existing methods of feature selection in scRNA-seq data analysis select features by deviance, or the most highly variable genes. In contrast, we sought to use the GFS algorithm to select deferentially distributed features that can help inform the underlying biology. We employed our algorithm in a multi-sample setting of relevant human memory CD4 T cell biology. We used an existing dataset of purified human memory CD4+ T cells, which were either left unstimulated (0 hr), or CD3/CD28 stimulated for 12 or 24 hours (12hr and 24hr, respectively). scRNA-seq was then performed on the cell populations at each of the three different time points \cite{SCPA}. In a dataset where the goal is to compare multiple time points concurrently to identify the features of interest, existing methods take a two-sample approach (comparing two time points at a time), which tends to decrease the power and sensitivity of the analysis. In GFS, we circumvent this problem by harnessing a multisample and multivariate graph based test as the basis.
	
	In order to have the recursion depth of our algorithm within control, we divided the set of features (genes) in the dataset into $8$ groups of equal size based on a pivot-based clustering approach. To elaborate, for every variable $i$ and every $\rho \in (0,1)$, we define $N_{\rho}(i)$ to be the set of all variables other than $i$, which have Spearman correlation at least $\rho$ with $i$, and select that variable $i$, for which the size of $N_\rho(i)$ is maximum, which we call a pivot. We then include the variables with highest correlation with the pivot $i$ in the first group. After constructing the first group, we leave it out subsequently, and apply the same mechanism on the remaining variables (by choosing a pivot from the left-out variables, and including in the second group the variables from among the remaining ones, whose correlation with this second step pivot is maximum). We continue this process till the eighth and final group has been constructed. We take $\rho = 0.5$ in our analysis, and if $N_{0.5}(i)$ happens to be empty for some pivot $i$, we reduce $\rho$ in decrements of $0.05$, till $N_{\rho}(i)$ becomes non-empty. Next, within each group, we carry out our MMCM test over $1000$ bootstrap iterations on randomly selected sub-samples of equal size from each class, in each of these iterations. Finally, we report the features that get selected across each of the $1000$ bootstrap iterations. 
	
	\begin{figure}
		\centering
		\includegraphics[width=0.6\linewidth]{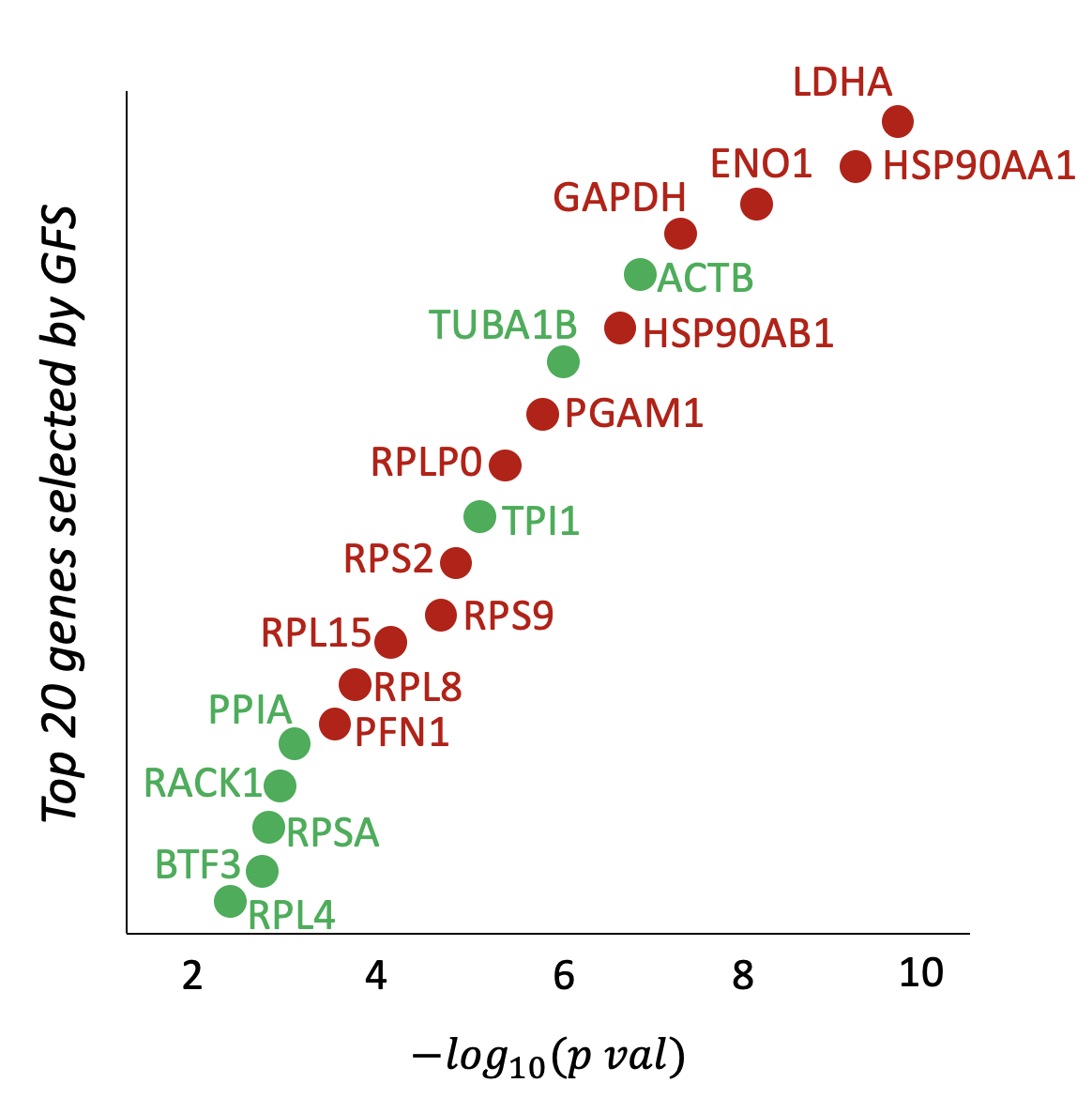}
		\caption{Top 20 differentially distributed genes between resting and activated human memory CD4 T cells. GFS detects known biologically important genes (in red) while also detecting new features of interest (genes labeled in green)}
	\end{figure}
	
	Utilizing the GFS approach to identify differentially distributed genes in the single cell transcriptomics setting, we discovered a variety of genes with both known and previously unrecognized roles in T cell biology (Fig. 6). For example, activated memory CD4 T cells rely on export of metabolites from the mitochondria to support epigenetic remodeling required for effector functions. Lactate dehydrogenase A (LDHA) in particular is known to be critical for glycolysis in CD4 T cells \cite{Tmetab, ldha}. Interestingly, glycolytic enzymes such as phosphoglycerate mutase-1 (PGAM1) and LDHA, as well as heat shock protein 90 (HSP90) molecules were found to be among the top 10 differentially distributed genes across the three subgroups based on GFS. Not only did our algorithm recapitulate aspects of known biology, but also highlighted previously unrecognized molecules such as receptor for activated c kinase 1 (RACK1) and basic transcription factor 3 (BTF3), which might play an important role in memory CD4 T cell activation.
	
	\subsection{Mice Protein Data}\label{sec:genomicdata}
	
	\subsubsection{Dataset Introduction} 
	\label{Dataset Introduction}
	
\begin{figure}[htbp]
		\centering
		\includegraphics[scale=0.45]{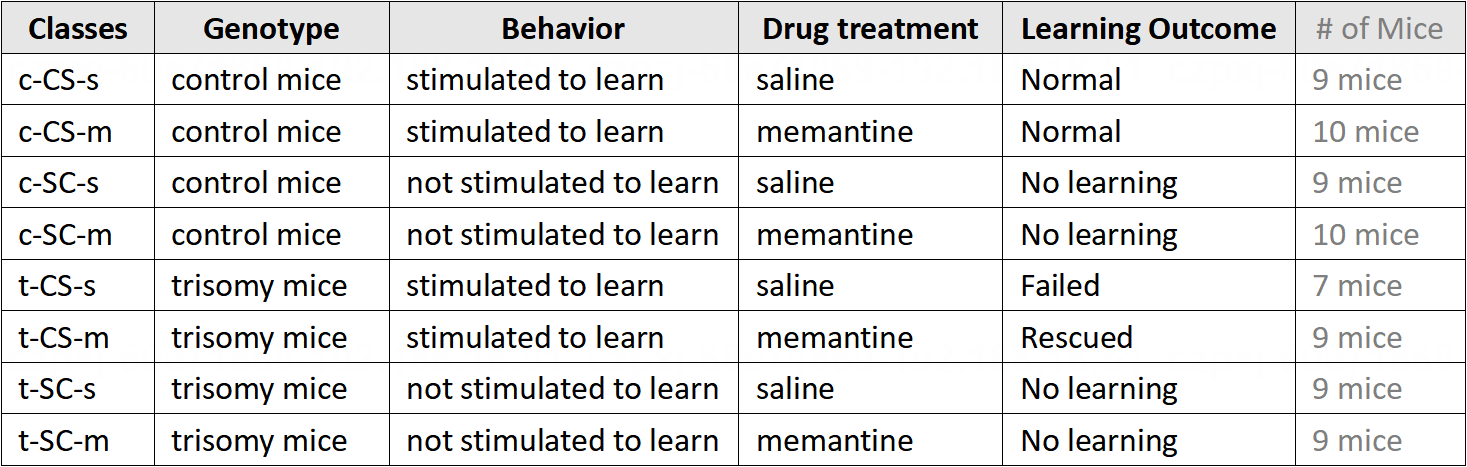}
		\caption{Classes of mice}
		\vspace{-13.5pt}
		\label{data_table}
	\end{figure}

	The \href{https://www.kaggle.com/datasets/muhammetvarl/mice-protein?datasetId=1040993}{\textit{Mice Protein dataset}} contains the expression level data of 77 protein markers in mice, which are generated from an experiment aiming to detect the important proteins that contribute to learning. As shown in Figure \ref{data_table}, the dataset is classified into 8 classes according to three aspects, namely the genotype of mice, stimulation behavior and drug treatment, which may lead to different learning outcomes. The dataset contains a total of 1080 measurements of protein response, each considered as an independent sample. 
	
	\subsubsection{Data Pre-processing} 
	\label{Data Pre-processing}
	
\begin{figure}[htbp]
		\centering
		\includegraphics[scale=0.4]{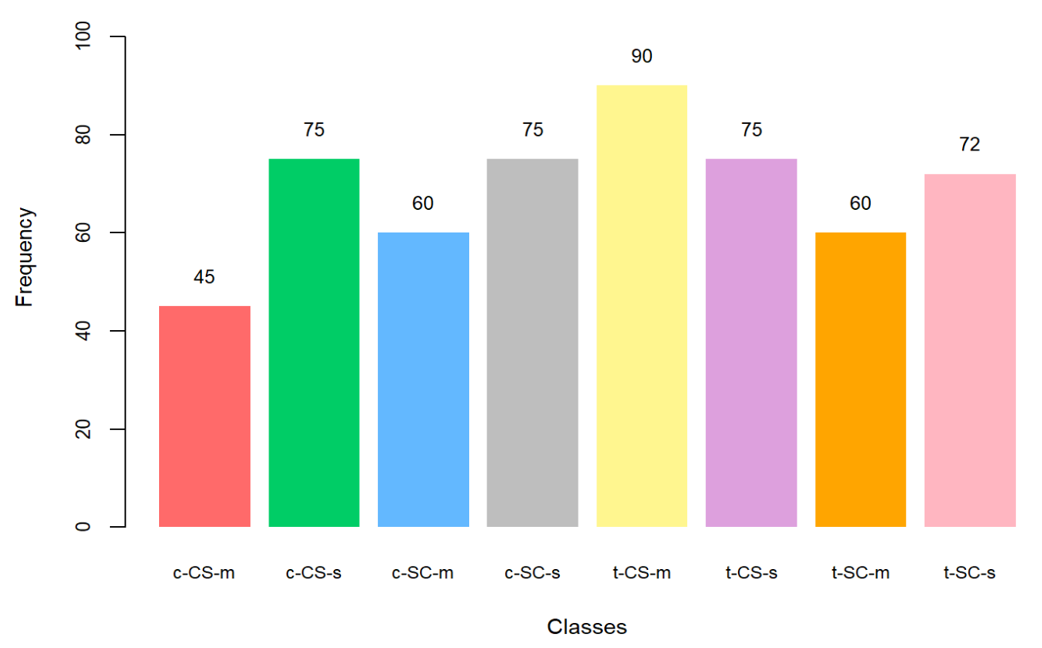}
		\caption{Frequency distribution of mice classes}
		\label{frequency_distribution}
	\end{figure}

	We first examined the basic data structure and filtered out 528 missing values. From Figure \ref{frequency_distribution}, we can see that the 8 classes of samples are unbalanced. 
	
\begin{figure}[htbp]
		\centering
		\includegraphics[scale=0.6]{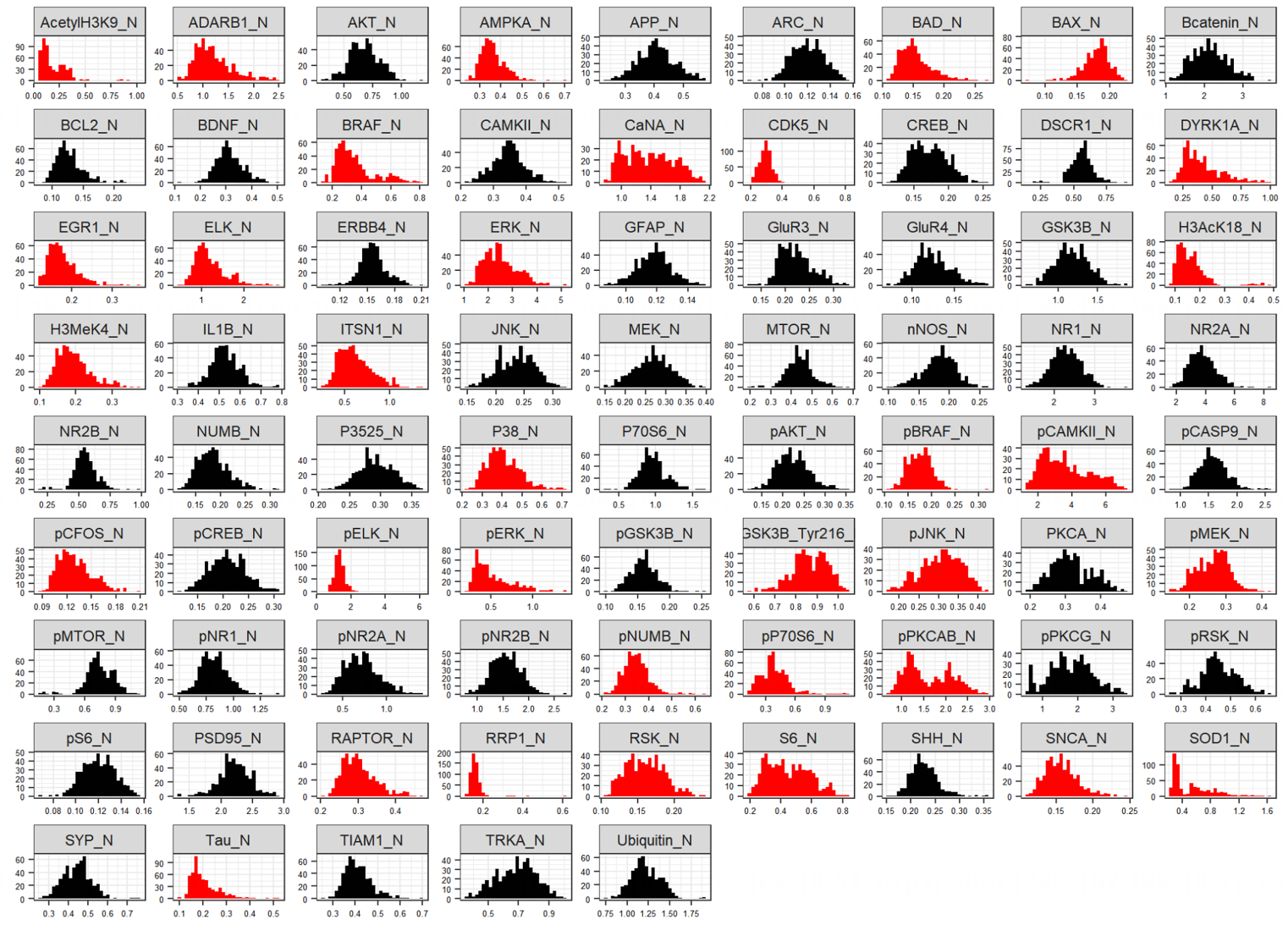}
		\caption{Feature distributions}
		\label{feature_distribution}
	\end{figure}

	Figure \ref{feature_distribution} shows that some features might not be normally distributed. However, since the MCM and MMCM tests are non-parametric and distribution-free, no distributional assumptions will be needed.
	
\begin{figure}[htbp]
		\centering
		\includegraphics[scale=0.8]{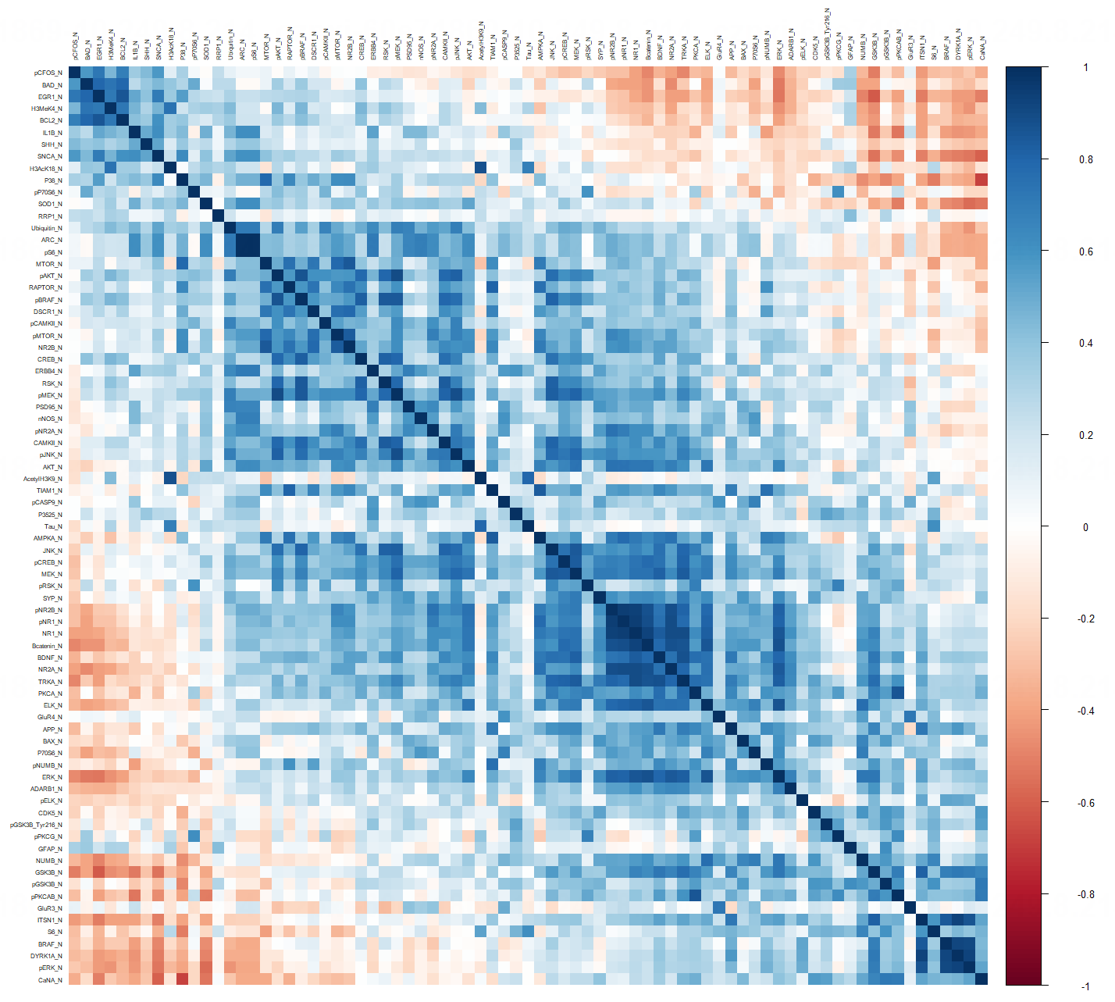}
		\caption{Correlation between features}
		\label{correlation}
	\end{figure}

	As Figure \ref{correlation} suggests, the majority of protein expression levels are highly correlated. To determine whether a significant difference exists between the 8 classes, we sought to find out the pivotal proteins that might be contributing to the overall difference, which in turn may help to detect the latent effective drug targets. 
	
	\subsubsection{Feature Selection Based on the MMCM Test}
	To detect the difference between 8 mice classes, we first applied the MMCM test, and the test statistic turns out to be 1679.85, much larger than $\chi^2_{0.95}(28)$ = 41.34. So the null hypothesis of no difference can be rejected at level $0.05$, thereby enabling us to conclude that there exists a statistical difference between 8 classes. 
	
\begin{figure}[htbp]
		\centering
		\includegraphics[scale=0.24]{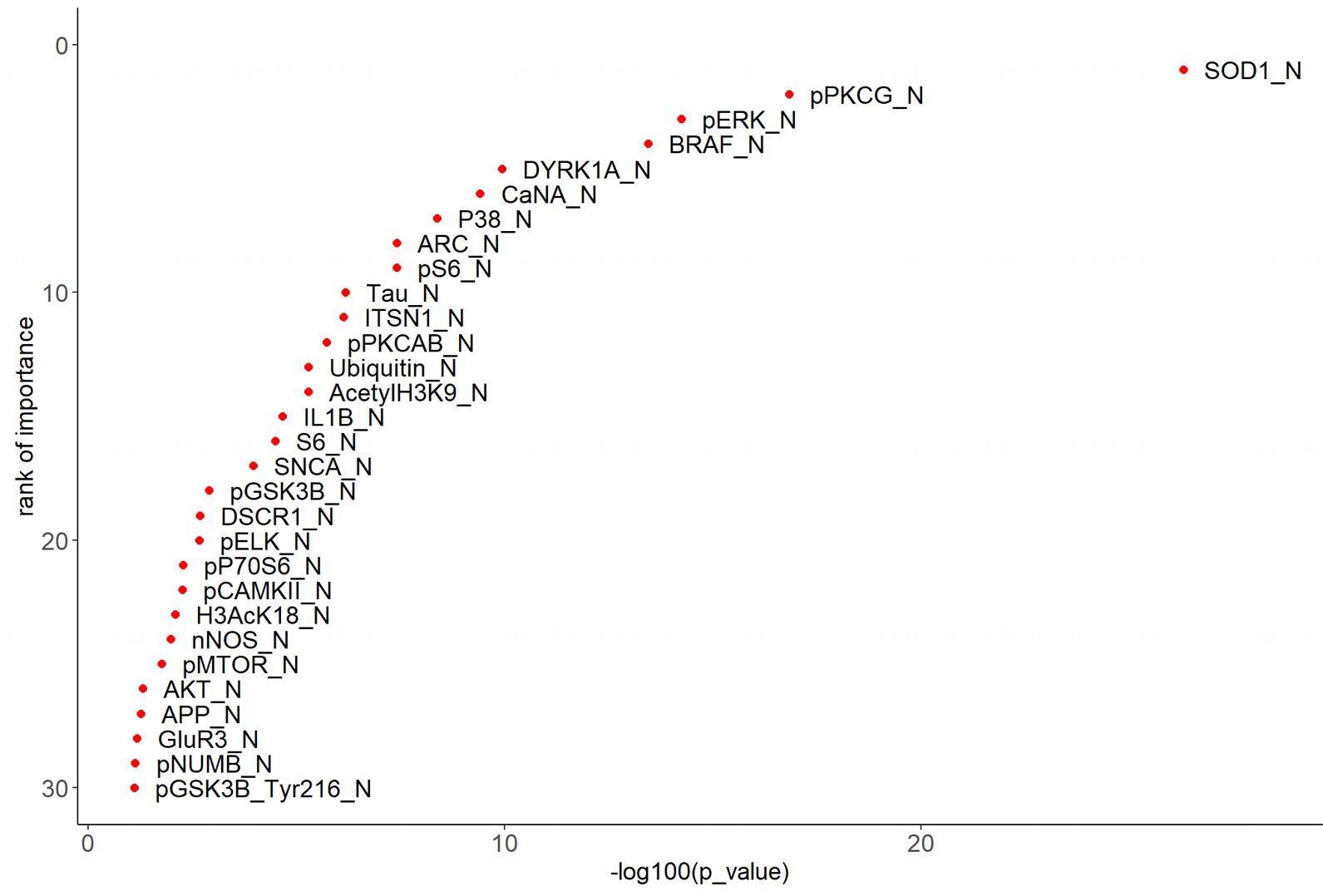}
		\caption{Top 30 proteins selected by GFS}
		\label{GFS}
	\end{figure}

	Based on the MMCM test result, we further employ the GFS algorithm to identify the most important proteins that contributed to the overall difference in learning outcomes of mice and also employ two traditional parametric methods for comparison.
	The GFS approach selects 47 significant proteins and Figure\ref{GFS} lists the top 30 based on the MMCM test p-values. 
	
	\subsubsection{Multinomial Logistic Regression with LASSO (MLR-LASSO)}
\begin{figure}[htbp]
		\centering
		\includegraphics[scale=0.65]{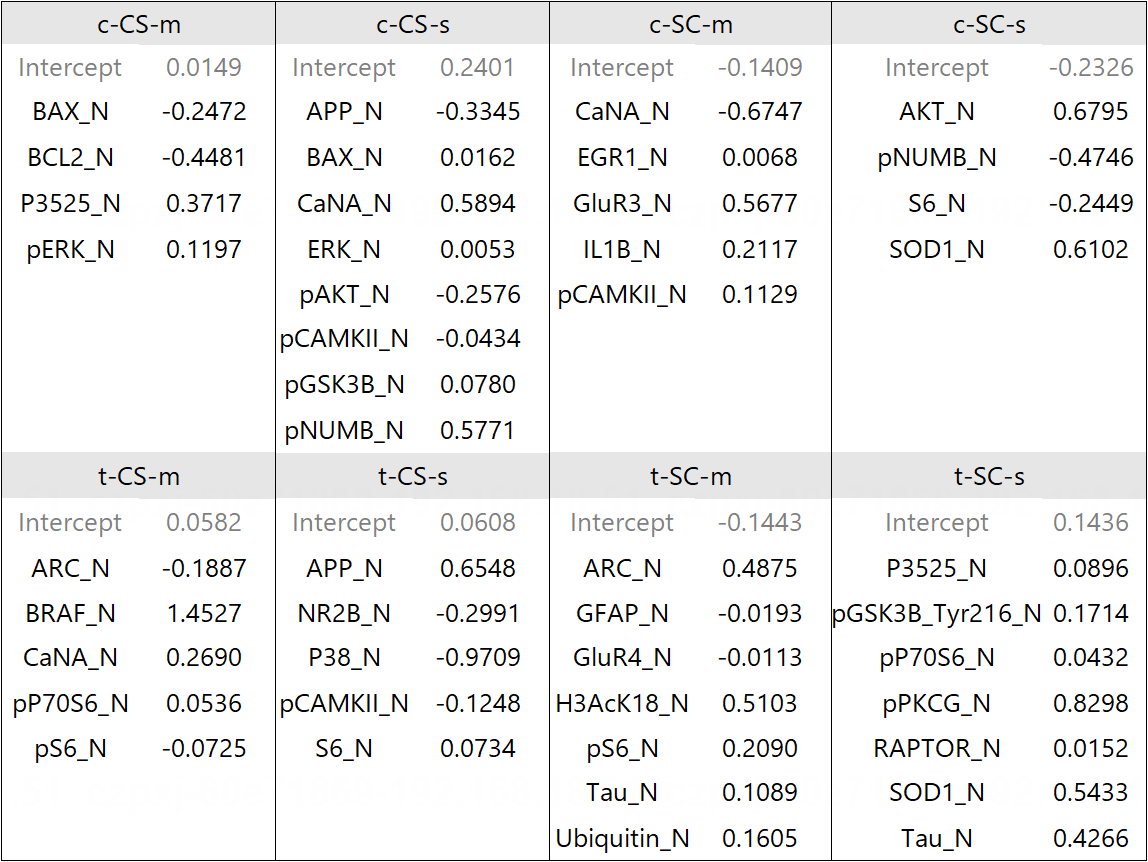}
		\caption{Proteins selected by Multinomial Regression with LASSO}
		\label{LASSO}
	\end{figure}

	We first standardized all the features based on their data distribution. Given the high correlation between most features, we incorporated the $L^1$ regularization method, namely LASSO (Least Absolute Shrinkage and Selection Operator), to establish a multinomial logistic regression model. LASSO is capable of both regularization and feature selection. This method gives 8 regression models on 8 classes and we subsequently select 31 important proteins in total as shown in Figure\ref{LASSO}. In addition, we test the model on $30\%$ samples and gain an accuracy of approximately $87.35\%$.
	
	\subsubsection{Random Forest (RF)}
\begin{figure}[htbp]
		\centering
		\includegraphics[scale=0.25]{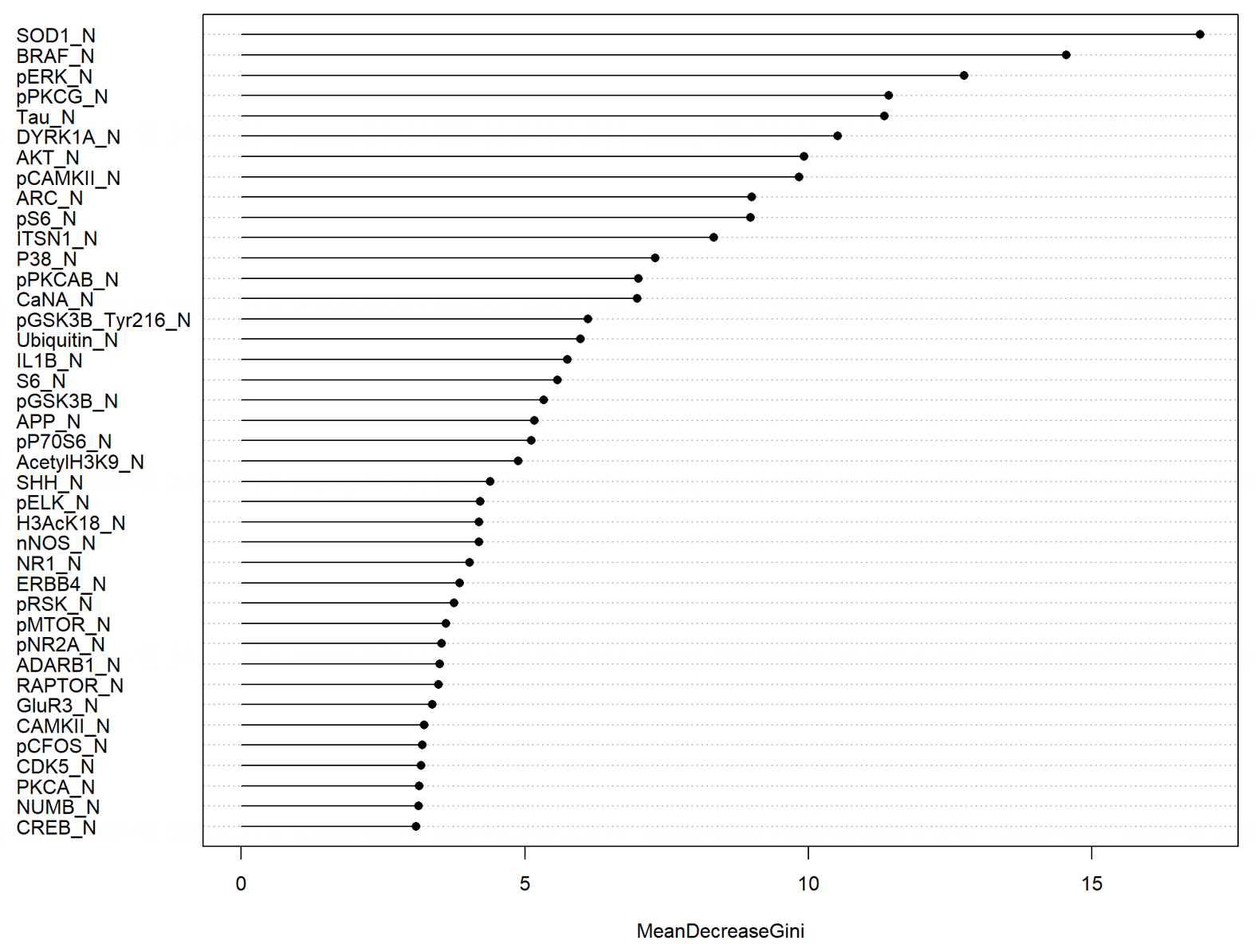}
		\caption{Feature importance of top 40 proteins}
		\label{RF}
	\end{figure}

	This approach gets the Feature Importance Score according to mean decrease in Gini coefficient and we take the top 40 important features in consideration as shown in Figure \ref{RF}. We also test this model on 30\% samples and achieve an accuracy of nearly 97.59\%. 
	
	\subsubsection{Result Analysis}
\begin{figure}[htbp]
		\centering
		\includegraphics[scale=1]{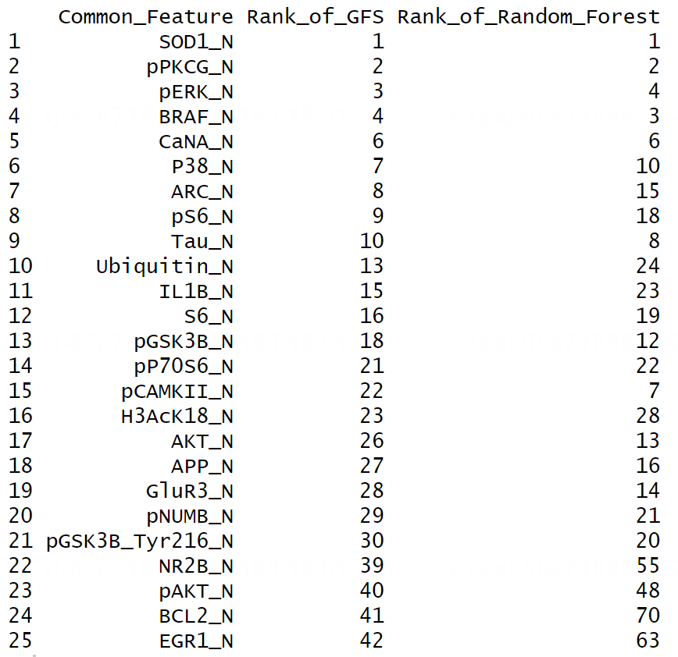}
		\caption{Common features selected by GFS and RF}
		\label{common}
	\end{figure}

\begin{figure}[htbp]
		\centering
		\includegraphics[scale=0.75]{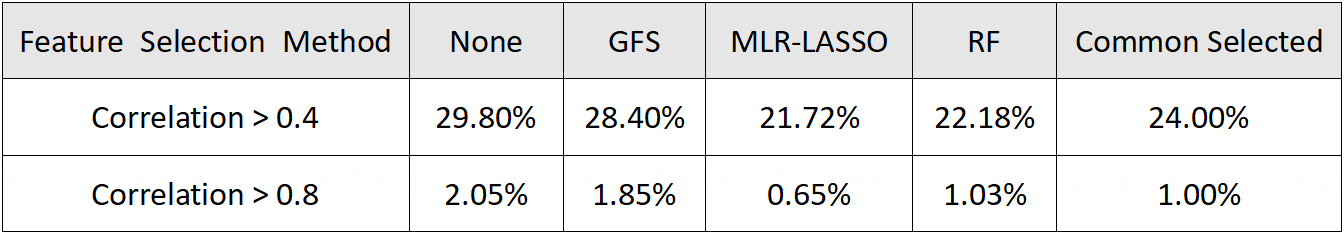}
		\caption{Proportion of highly related feature pairs}
		\label{cor_comparison}
	\end{figure}

	The common important features (proteins) selected by the above three methods are listed in Figure \ref{common}. As shown in Figure \ref{cor_comparison}, the highly correlated feature pairs in the original dataset would be reduced by employing the feature selection methods. The common selected proteins are expected to be the most important features. Furthermore, according to the rank of importance given by both the GFS and the Random Forest method (see Figure \ref{common}), proteins ranked highly by both methods can be surmised as more effective drug targets. 
	
	\subsection{Brain Cancer Data}
	\subsubsection{Dataset Introduction}
\begin{figure}[htbp]
		\centering
		\includegraphics[scale=0.6]{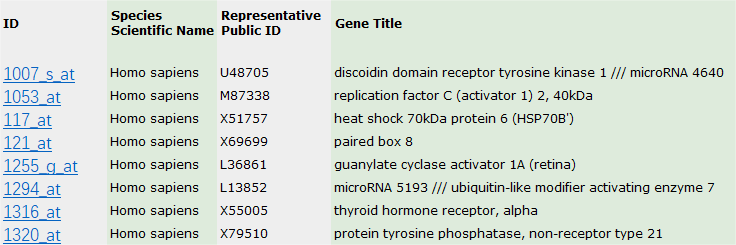}
		\caption{Description of gene probes in the brain cancer dataset}
		\label{brain_description}
	\end{figure}

	{\textit{The Brain Cancer dataset}} is publicly available from the Gene Expression Omnibus (GEO) data base \cite{b35}, and comprises of a transcriptomics study to explore the underlying relationship between various genes and the difference among types of brain tumors. The gene expression profiles of 130 surgical tumor and normal brain samples were obtained via Affymetrix HG-U133plus2 chips. To isolate the gene expression of pivotal immune cell markers, the gene expression profiles of various types of brain tumors and normal brains were filtered. Nearly 56,000 genes were profiled in total and all samples were labeled in 5 classes, namely ependymoma, glioblastoma, medulloblastoma, normal, and pilocytic astrocytoma. An example of some of the genes included with corresponding representative gene IDs and titles are shown in Figure \ref{brain_description}.
	
	\subsubsection{Data Pre-Processing}
	Considering there are over 56,000 genes in the brain cancer dataset, applying our GFS method would be inherently challenging due to computational resources. Hence we pre-selected highly variable genes based on their expression level by using Median Absolute Difference (MAD):
	$$\text{MAD}=median_i (median_j |S_{i,j}-\textbf{S}_i|)$$
	where $S_{ij},\,i=1,\cdots,56000,\,j=1,\cdots,130$ are values in the $(i,j)$-th entries in the gene expression matrix and $\textbf{S}_i$ is the median of $(S_{i,j})_{j\ge 1}$. We re-ordered all genes and selected the first 5000 with the largest MAD for downstream analyses.
	
	\subsubsection{Structure of dataset}
\begin{figure}[htbp]
		\begin{minipage}[t]{0.5\linewidth}
			\centering
			\includegraphics[width=\textwidth]{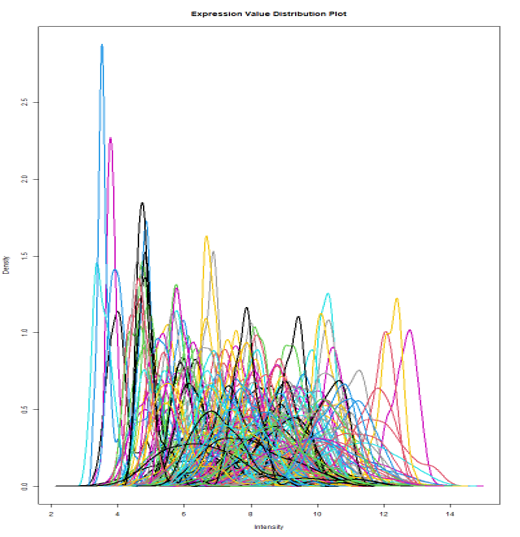}
			\centerline{(a) Density for genes}
		\end{minipage}%
		\begin{minipage}[t]{0.5\linewidth}
			\centering
			\includegraphics[width=\textwidth]{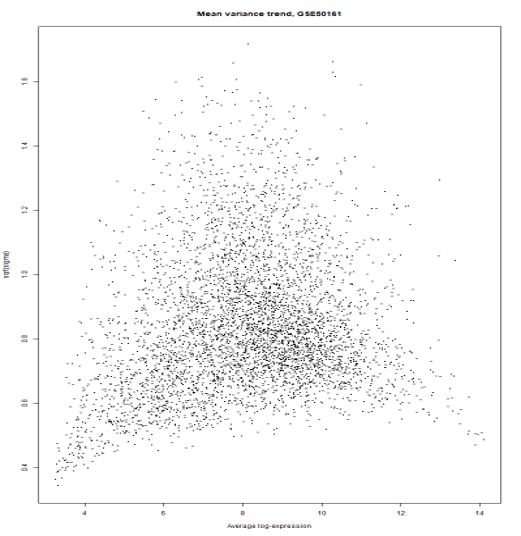}
			\centerline{(b) Mean-Variance Trend}
		\end{minipage}
		\caption{EDA on Brain Cancer dataset}
		\label{htbp1}
	\end{figure}

	After pre-selecting genes, we performed exploratory data analysis. The left window of Figure \ref{htbp1} shows the expression value distribution plot for 5,000 genes. These distributions have clear dissimilarities in terms of means and variances, as is revealed by the plots. The right window of Figure \ref{htbp1} shows the mean and variance trend of all the 5,000 variables. These data points show a clear pattern, thereby suggesting a quantifiable biological structure, which we expect to be picked up in our feature selection algorithm. 
	
	\subsubsection{Pre-Clustering}
	Genes can be co-expressed as part of a shared chemical or biological function. To capture this, we decided to apply Weighted Gene Co-expression Network Analysis (WGCNA)\cite{b36}. It is a widely used systems biology method for constructing a co-expression network based on the expression matrix of various genes, and allowed us to cluster genes into different key modules. By applying this method, we divided our 5,000 genes into 23 modules where each module corresponds to a particular biological pathway. 
	
	\subsubsection{Results Summary}
\begin{figure}[htbp]
		\centering
		\includegraphics[scale=0.8]{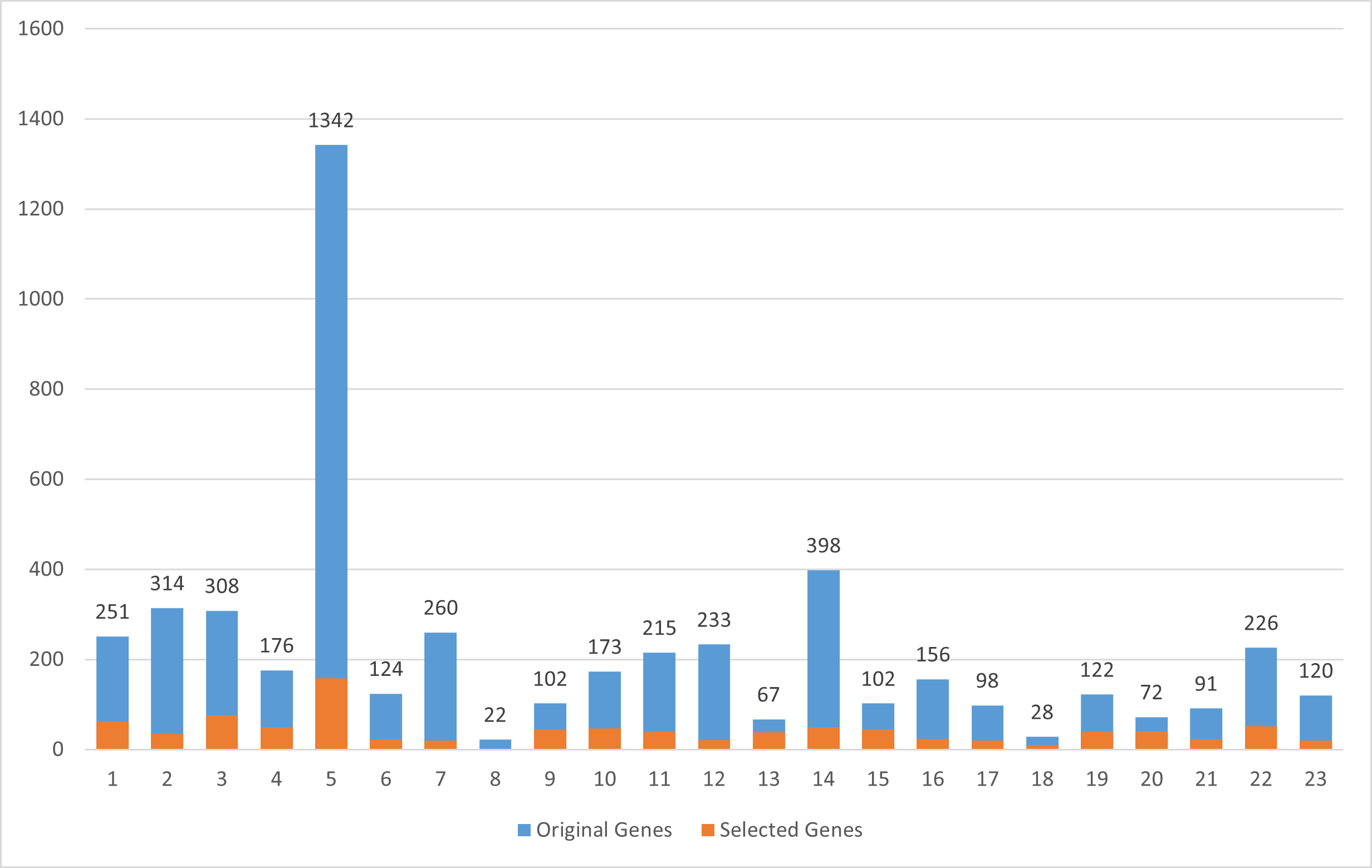}
		\caption{Bar chart for selected variables}
		\label{Bar chart for Selected variables}
	\end{figure}

	We applied the MMCM test to the $130\times5,000$ data matrix. We observed that the MMCM test statistic was much greater than $t_{8385}(0.05)$, indicating that the differences among various types of brain tumors were significant. Next, we applied our GFS method to each of the 23 modules to select the most significant variables within them. We set the significance level to 0.05 and selected a total of 935 variables. The results of this analysis are illustrated in Figure \ref{Bar chart for Selected variables}. From the bar chart, we can see that in Module 8, we did not select any features from the 22 genes included in that module. This could be due to our choice of significance level, or it could be because Module 8 has unique characteristics that make it less informative for identifying significant variables.
	
\begin{figure}[htbp]
		\centering
		\includegraphics[scale=0.6]{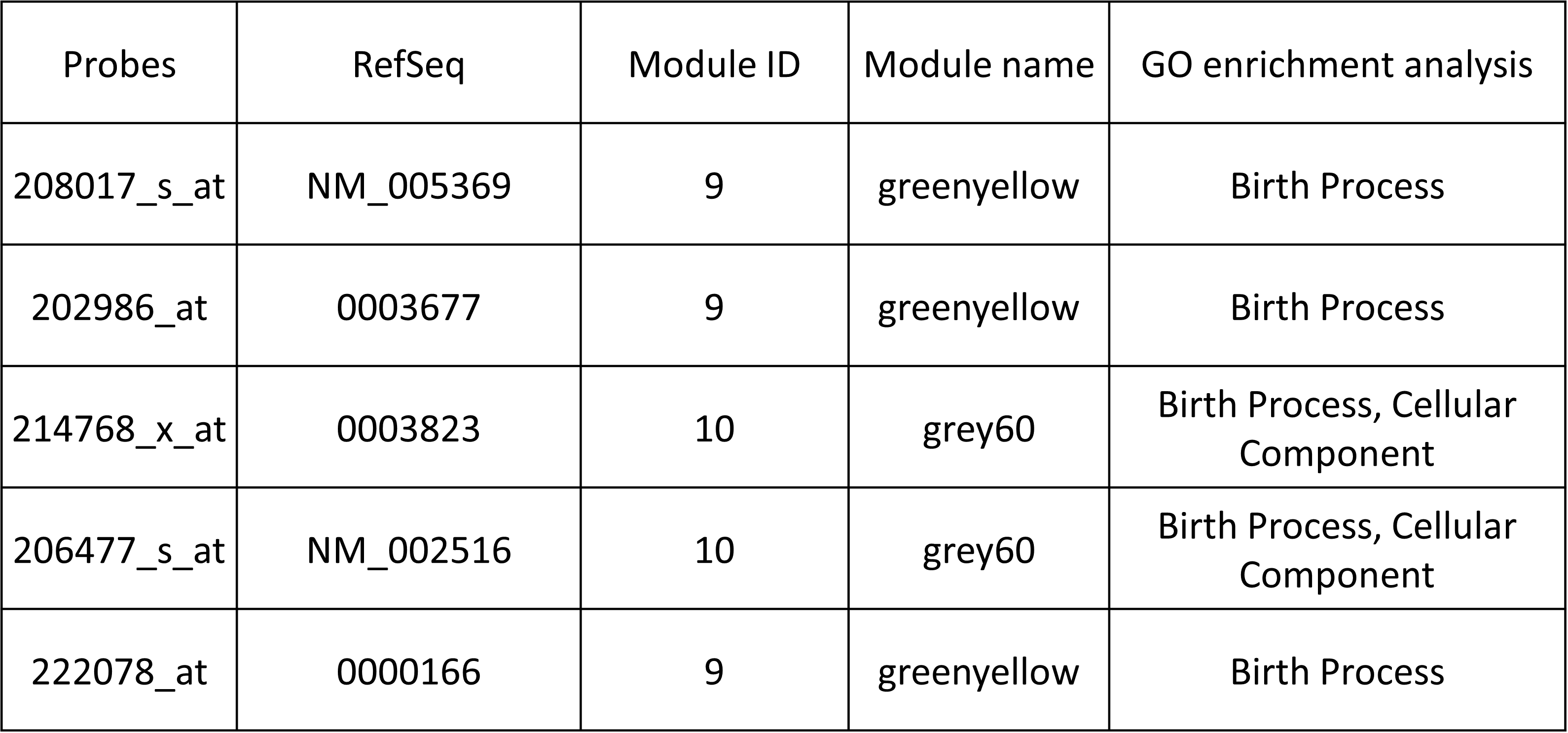}
		\caption{Table for top 5 genes}
		\label{Table for top 5 genes}
	\end{figure}

	To gain further insight into the relationship between types of brain tumors and our selected genes, we sorted the variables by their corresponding adjusted $p$-values, from small to large. This approach allows us to focus on genes that have a high degree of confidence in differentiating between tumor types and have a greater contribution during cellular processes. Figure \ref{Table for top 5 genes} shows the first 5 genes that were sorted. Interestingly, all of these genes belong to Modules 9 and 10, which are likely related to brain tumors, as confirmed by Gene Ontology (GO) enrichment analysis \cite{b37}, a method for identifying over-represented biological functions or processes associated with a set of genes or proteins of interest. This analysis suggests that most genes in these modules have functions related to birth processes (germline mutations), which are important in the development of brain tumors.
	
	\begin{figure}[htbp]
		\centering
		\includegraphics[scale=0.6]{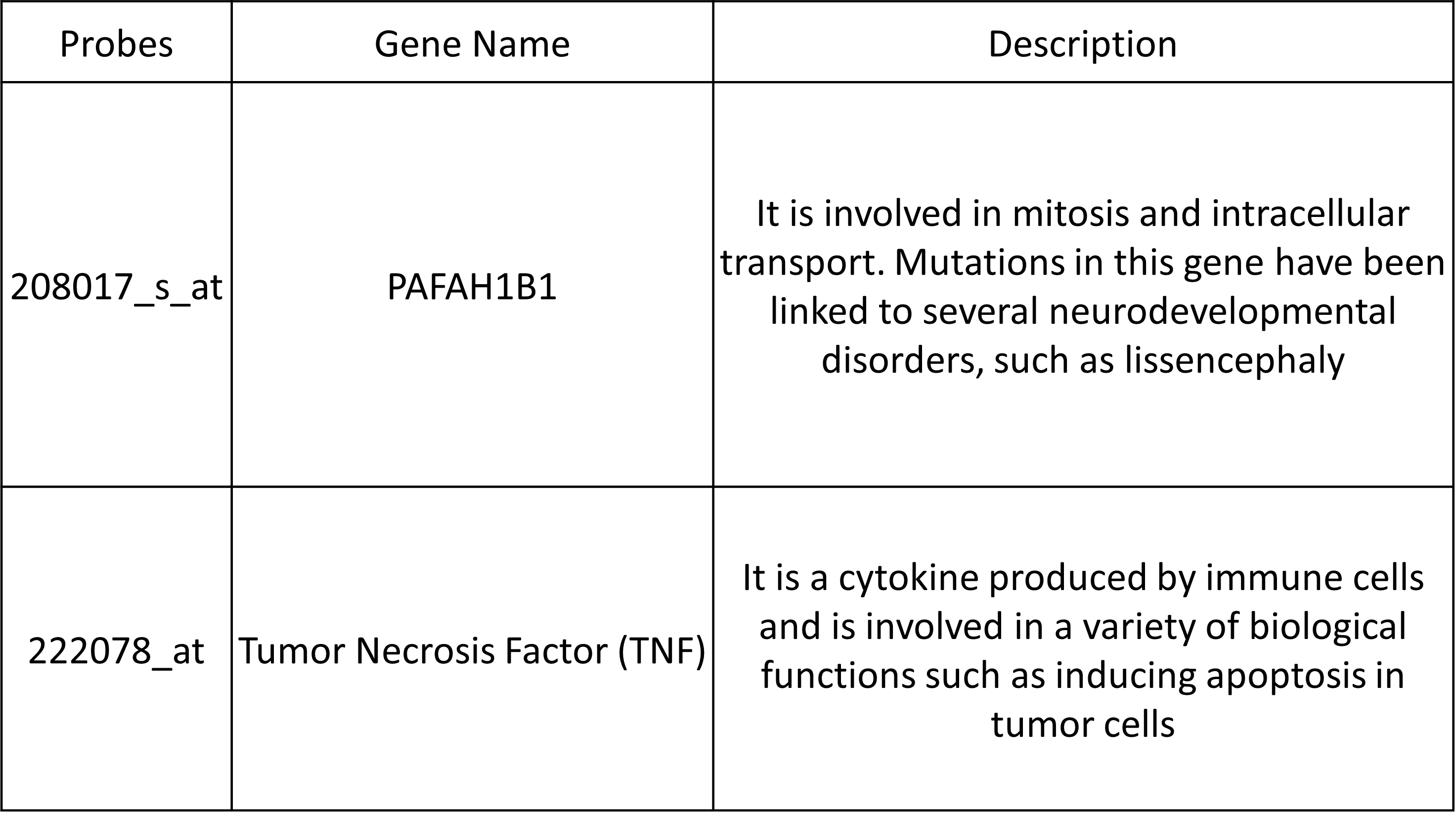}
		\caption{Formal description of original genes}
		\label{Formal description of original genes}
	\end{figure}
	
	\begin{figure}[htbp]
		\centering
		\includegraphics[scale=0.6]{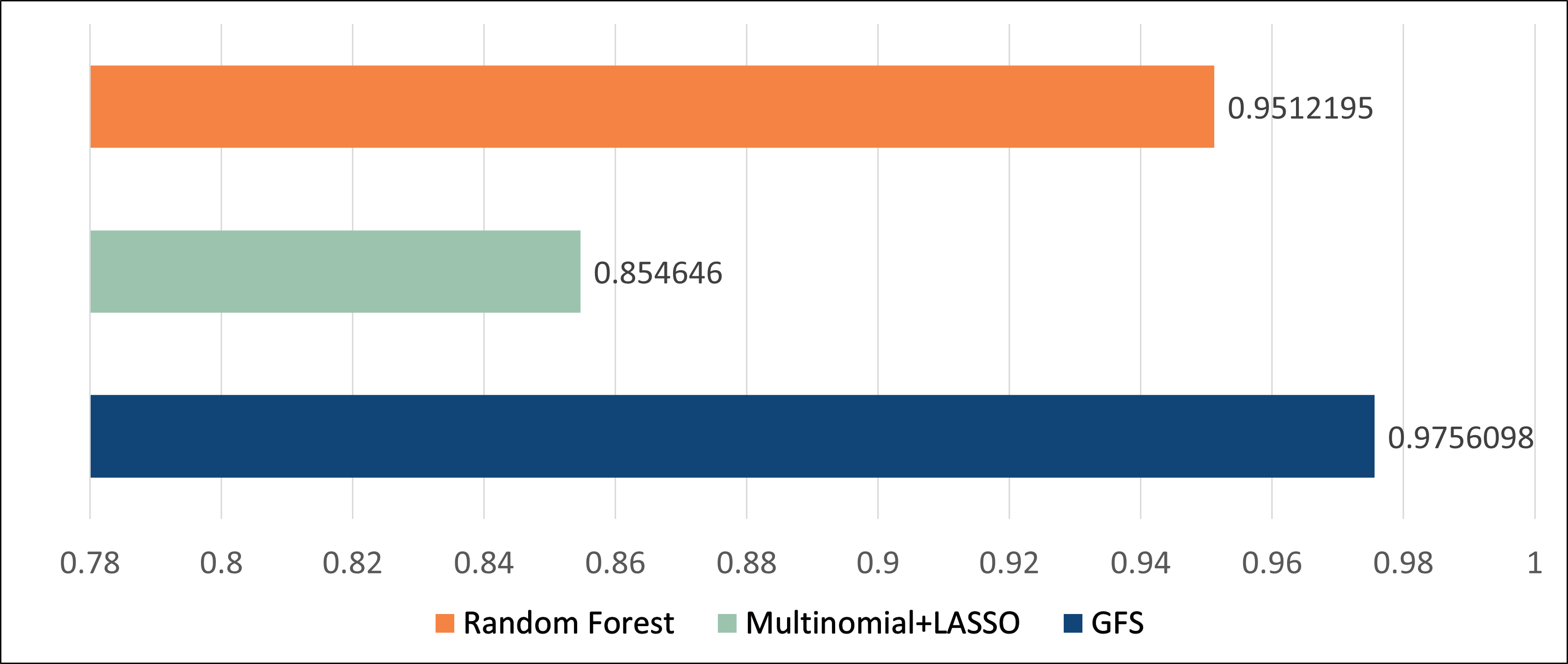}
		\caption{Comparison between common feature selection methods}
		\label{Comparison between common feature selection methods}
	\end{figure}
	To evaluate the performance of our GFS method, we compared it with other common feature selection methods, namely Multinomial regression based on LASSO and Random Forest. We use SVM with RBF kernel as our classification model and apply it to the selected features. We divide our samples into training and test datasets with a proportion of 0.7 for the training dataset. During the training procedure, we perform a grid search to find the best hyperparameters to avoid overfitting.
	The results of our comparison are shown in Figure \ref{Comparison between common feature selection methods}. The GFS method outperforms other feature selection methods in terms of validation accuracy, thereby suggesting that the combination of the MMCM test and the GFS algorithm might serve as a more efficient and applicable approach.
	
	Overall, our study provides a promising approach for identifying significant genes in brain tumor classification and demonstrates the effectiveness of our GFS method in feature selection for classification models.
	
	\section{Discussion}\label{sec:discussion}
	In this paper, we describe a graph-based hierarchical clustering approach for multi-dimensional feature selection under a complete nonparametric multisample setup. This method, which we call the GFS algorithm, is consistent, i.e., it reports all the true features contributing to the distributional difference with high probability. Concurrently, GFS guarantees a control on the probability of selecting at least one non-contributing feature. Our simulations show that this graph-based approach provides a substantial improvement over classical parametric tests for location shift, such as the Kruskal Wallis test, in the generality. Further still, through analysis of the Portuguese forest fires data, a single cell transcriptomics data, a mice protein data and a brain cancer data, we highlight the power of GFS at unearthing and capturing variables that can provide deeper insights into the process being examined. With growing popularity of scRNA-seq data in biomedical research, we envision that our method can find a unique place in instances wherein multiple cell types of patient cohorts are being compared, and existing differential gene expression methods find limited utility.
	
	At the heart of our feature-selection method is the nonparametric multisample graph-based testing approach. We would like to emphasize that the hierarchical clustering procedure is just a tool, and can be replaced by other suitable alternatives. For example, a potentially promising alternative may be to implement a ``bootstrap" type of algorithm, where the basic idea is to select a certain number of random subsets of a fixed cardinality from the feature set, and carry out the multisample graph-based test on the sub-datasets obtained by truncating the original dataset at those random feature subsets. All the significant sub-datasets are then pooled together keeping track of the multiplicity of each feature appearing in the pooled multiset. The variables in this pooled multiset can then be reported as significant features, in decreasing order of their multiplicities. Establishing theoretical guarantees for the bootstrap approach may be an interesting direction for future research. Nonetheless, in this work we establish that multivariate graph-based methods can be powerful tools for developing feature selection algorithms that detect meaningful features which might not be captured by existing univariate approaches.
	
	\section{Acknowledgment}
	S.G. was supported by the National Science Foundation BIGDATA grant IIS-1837931. S.M. was supported by the National University of Singapore start-up grant WBS
	A0008523-00-00 and the Faculty of Science Tier 1 grant WBS A-8001449-00-00.
	We are immensely grateful to Nancy Zhang for suggesting us this problem and giving some nice ideas to attack it, including the alternative bootstrap-type approach. We would also like to thank Bhaswar Bhattacharya for several helpful discussions throughout the preparation of this manuscript.
	
	\section{Appendix}  
	In this section, we prove some technical lemmas necessary for the proof of Theorem \ref{mainthm}, and show exact numerical results for some of the simulations we did earlier.
	\subsection{Technical Results}
	
	In this section, we prove some technical results needed to prove Theorem \ref{mainthm}.
	
	\begin{lem}\label{unique1}
		A signal set is unique.
	\end{lem}
	\begin{proof}
		Suppose that $L_1$ and $L_2$ are two signal sets. It is enough to show that $L_1 \subseteq L_2$, since the other inclusion will then follow by symmetry. Towards showing this, take $i \in L_1$. By (\ref{cc2}), $H_{0,\{i\}}$ is not true, i.e. there exist $1\le u<v\le K$ such that $F_{u[\{i\}]} \ne F_{v[\{i\}]}$. This means that for all $L \supseteq \{i\}$, $F_{u[L]} \ne F_{v[L]}$, and hence, $H_{0,L}$ is not true, and therefore, by (\ref{cc1}), $L \ne L_2^c$. We conclude that $i \notin L_2^c$, i.e. $i \in L_2$. This completes the proof.
	\end{proof}
	
	\begin{lem}\label{ap2}
		Let $\mathcal{T}_0 := \{C\in \mathcal{T}: H_{0,C} ~\textrm{is true}\}$, and let $\mathcal{T}_{\mathrm{rej}} := \{C\in \mathcal{T}: \max_{U \in \mathcal{T}: U \supseteq C} p_{\mathrm{adj}}^{U} \le \alpha\}$. Then, $$\p\left(\mathcal{T}_0 \bigcap \mathcal{T}_{\mathrm{rej}} \ne \emptyset\right) \le \alpha~.$$
	\end{lem}
	
	\begin{proof}
		The proof of Lemma \ref{ap2} is exactly similar to the proof of Theorem 1 in \cite{hierarchical}, but since the framework there is a regression setup, we replicate their argument and show that everything remains unchanged even in our nonparametric setup. To start with, let us define:
		$$\mathcal{T}_0^{\max} := \left\{C \in \mathcal{T}_0:~\nexists~ V \in \mathcal{T}_0~\textrm{with}~V \supseteq C\right\}~.$$
		Now, it is evident that:
		$$\left\{\mathcal{T}_0 \bigcap \mathcal{T}_{\mathrm{rej}} \ne \emptyset\right\} = \left\{\mathcal{T}_0^{\max} \bigcap \mathcal{T}_{\mathrm{rej}} \ne \emptyset\right\} \subseteq \left\{\mathcal{T}_0^{\max} \bigcap \{C \in \mathcal{T}: p_{\mathrm{adj}}^{C} \le \alpha\} \ne \emptyset\right\}~.$$
		Hence, we have:
		$$\p\left(\mathcal{T}_0 \bigcap \mathcal{T}_{\mathrm{rej}} \ne \emptyset\right) \le \sum_{C\in \mathcal{T}_0^{\max}} \p(p_{\mathrm{adj}}^{C} \le \alpha) = \sum_{C\in \mathcal{T}_0^{\max}} \frac{\alpha |C|}{d}~.$$
		Since any two distinct elements of $\mathcal{T}_0^{\max}$ are disjoint, it follows that: $$\sum_{C\in \mathcal{T}_0^{\max}} |C| \le d~.$$ Hence, we have:
		$$\p\left(\mathcal{T}_0 \bigcap \mathcal{T}_{\mathrm{rej}} \ne \emptyset\right) \le \alpha~,$$ completing the proof of Lemma \ref{ap2}.
	\end{proof}
	
	\subsection{Numerics for location and scale families :}\label{num11}
	In this section, we tabulate the exact numerical results for the simulations in Sections \ref{loc1} and \ref{sc1}.
	
	\begin{table}[h]
		\centering
		\begin{tabular}{|c||c|c|c|}
			\hline
			$\theta\downarrow$ &  FWER & FDR & Power \\  
			\hline
			\hline
			\multirow{1}{*}{.15} & 0.45 & 0.08 & 0.33 \\
			\hline
			\multirow{1}{*}{.20} & 0.04 & 0.01 & 0.31 \\ 
			\hline
			\multirow{1}{*}{.25} & 0.07 & 0.02 & 0.35 \\ 
			\hline
			\multirow{1}{*}{.30} & 0.04 & 0.005 & 0.41 \\ 
			\hline
			\multirow{1}{*}{.35} & 0.03 & 0.003 & 0.51 \\ 
			\hline
			\multirow{1}{*}{.40} & 0.04 & 0.003 & 0.80 \\ 
			\hline
			\multirow{1}{*}{.45} & 0.05 & 0.003 & 0.97 \\ 
			\hline
			\multirow{1}{*}{.50}& 0.03 & 0.001 & 1.00 \\ 
			\hline
			\multirow{1}{*}{.55} & 0.03 & 0.004 & 1.00 \\ 
			\hline
			\multirow{1}{*}{.60} & 0.03 & 0.001 & 1.00 \\ 
			\hline
			\multirow{1}{*}{.65} &0.02 & 0.004 & 1.00 \\ 
			\hline
			\multirow{1}{*}{.70} & 0.02 & 0.004 & 1.00 \\ 
			\hline
			\multirow{1}{*}{.75} &0.03 & 0.006 & 1.00 \\ 
			\hline
		\end{tabular}
		
		\caption{Location Setting: FWER, FDR and Power}
		\label{table:normallocation}
	\end{table}
	
	\begin{table}[h]
		\centering
		\begin{tabular}{|c||c|c|c|}
			\hline
			$\theta\downarrow$ &  FWER & FDR & Power \\  
			\hline
			\hline
			\multirow{1}{*}{1.0} &  1.00 & 0.58 & 0.38\\
			\hline
			\multirow{1}{*}{1.5} & 0.99 & 0.48 & 0.44 \\
			\hline
			\multirow{1}{*}{2.0} & 0.98 & 0.44 & 0.48 \\
			\hline
			\multirow{1}{*}{2.5} & 0.98 & 0.44 & 0.52  \\
			\hline
			\multirow{1}{*}{3.0} &  0.99 & 0.38 & 0.57\\
			\hline
			\multirow{1}{*}{3.5} & 0.98 & 0.34 & 0.64 \\
			\hline
			\multirow{1}{*}{4.0} & 0.97 & 0.30 & 0.68 \\
			\hline
			\multirow{1}{*}{4.5} & 0.96 & 0.25 & 0.76  \\
			\hline
			\multirow{1}{*}{5.0} &  0.92 & 0.17 & 0.79\\
			\hline
			\multirow{1}{*}{5.5}& 0.92 & 0.17 & 0.84 \\
			\hline
			\multirow{1}{*}{6.0} & 0.77 & 0.12 & 0.89 \\
			\hline
			\multirow{1}{*}{6.5} & 0.76 & 0.10 & 0.91  \\
			\hline
			\multirow{1}{*}{7.0} &  0.6 & 0.08 & 0.93\\
			\hline
			\multirow{1}{*}{7.5} & 0.54 & 0.05& 0.95 \\
			\hline
			\multirow{1}{*}{8.0} & 0.41 & 0.03 & 0.96 \\
			\hline
			\multirow{1}{*}{8.5} & 0.40 & 0.03 & 0.977 \\
			\hline
			\multirow{1}{*}{9.0} & 0.32 & 0.02 & 0.982 \\
			\hline
			\multirow{1}{*}{9.5} & 0.16 & 0.01 & 0.986  \\
			\hline
			\multirow{1}{*}{10.0} &  0.18 & 0.02 & 0.988\\
			\hline
			\multirow{1}{*}{10.5} & 0.15 & 0.01 & 0.994 \\
			\hline
			\multirow{1}{*}{11.0} & 0.09 & 0.004 & 0.995 \\
			\hline
			\multirow{1}{*}{11.5} & 0.07 & 0.006 & 0.997  \\
			\hline
			\multirow{1}{*}{12.0} &  0.04 & 0.003
			& 0.996\\
			\hline
			\multirow{1}{*}{12.5}& 0.06 & 0.006 & 0.995 \\
			\hline
			\multirow{1}{*}{13.0} & 0.06 & 0.006 & 0.998\\
			\hline
			\multirow{1}{*}{13.5} & 0.03 & 0.001 & 0.999  \\
			\hline
			\multirow{1}{*}{14.0} &  0.04 & 0.004 & 0.999\\
			\hline
			\multirow{1}{*}{14.5} & 0.04 & 0.002 & 0.999 \\
			\hline
			\multirow{1}{*}{15.0} & 0.05 & 0.004 & 1 \\
			\hline
		\end{tabular}
		
		\caption{Scale Setting: FWER, FDR and Power}
		\label{table:normalscale}
	\end{table}


\bigskip
	\newpage
	\begin{thebibliography}{99}
		\bibitem{bickel}
		P. J. Bickel, A distribution free version of the Smirnov two sample test in the $p$-variate case, {\it Annals of Mathematical Statistics}, Vol. 40, 1--23, 1969.
		
		\bibitem{biswas}
		M.~Biswas, M.~Mukhopadhyay, and A.~K. Ghosh,
		\newblock {A distribution-free two-sample run test applicable to
			high-dimensional data}, 
		\newblock {\em Biometrika}, 101(4):913--926, 10 2014.
		
		\bibitem{fr}
		J.~H. Friedman and L.~C. Rafsky,
		\newblock Multivariate generalizations of the Wald-Wolfowitz and Smirnov two-sample tests, 
		\newblock {\em Ann. Statist.}, 7(4):697--717, 07 1979.
		
		\bibitem{morteza}
		M.H. Chehreghani, 
		\newblock Hierarchical Correlation Clustering and Tree
		Preserving Embedding,
		\newblock {arXiv:2002.07756}, 2020.
		
		\bibitem{gc}
		J. D. Gibbons and S. Chakraborty, {\it Nonparametric Statistical Inference}, Fourth Edition. Marcel Dekker Inc., 2003. 
		
		
		\bibitem{hastie}
		T. Hastie, R. Tibsirani, J. Friedman,
		\newblock The Elements of Statistical Learning; Data Mining, Inference and Prediction.
		\newblock {NewYork, Springer}, 2001.
		
		
		\bibitem{ks}
		W.~H. Kruskal.
		\newblock A nonparametric test for the several sample problem, 
		\newblock {\em Ann. Math. Statist.}, 23(4):525--540, 12 1952.
		
		\bibitem{kswallis}
		W.~H. Kruskal and W.~A. Wallis, 
		\newblock Use of ranks in one-criterion variance analysis, 
		\newblock {\em Journal of the American Statistical Association},
		47(260):583--621, 1952.
		
		\bibitem{mann_whitney}
		H. B. Mann and D. R. Whitney, 
		\newblock On a test of whether one of two random variables is stochastically larger than the other,
		\newblock{\em Annals of Mathematical Statistics}, Vol. 18(1), 50--60, 1947. 
		
		\bibitem{hcscheme}
		S. C. Johnson,
		Hierarchical clustering schemes
		\newblock{\em Psychometrika}, Vol. 32(3), 241--254, 
		1967.
		
		\bibitem{hcjspi}
		D. Zhu, D. P. Guralnik, X. Wang, X. Li and B. Moran Statistical properties of the single linkage hierarchical clustering estimator. 
		\newblock{\em Journal of Statistical Planning and Inference}, Vol. 185, 15--28, 2017.
		
		\bibitem{hierarchical}
		N. Meinshausen.
		\newblock Hierarchical testing of variable importance,
		\newblock {\em Biometrika}, 95, 2, 265-278, 2008.
		
		\bibitem{treeform}
		Daniel Müllner.
		\newblock Modern hierarchical, agglomerative clustering algorithms,
		\newblock {arXiv:1109.2378},2011
		
		\bibitem{mukherjee}
		S. Mukherjee, D. Agarwal, N. Zhang and B.B. Bhattacharya,
		\newblock Distribution-free Multisample Test Based on Optimal Matching with Applications to Single Cell Genomics,
		\newblock {\em Journal of the American Statistical Association}, \url{https://doi.org/10.1080/01621459.2020.1791131}
		
		
		\bibitem{rosen}
		P.~R. Rosenbaum.
		\newblock An exact distribution-free test comparing two multivariate
		distributions based on adjacency,
		\newblock {\em Journal of the Royal Statistical Society B}, 67:515--530, 2005.
		
		\bibitem{bioinform}
		Y. Saeys, I. Inza, P. Larra\~naga,
		\newblock A review of feature selection techniques in bioinformatics. 
		\newblock {\em Bioinformatics}, 23: 2507-–2517, 2007.
		
		\bibitem{SCPA}
		J.~A. Bibby, D. Agarwal D, T. Freiwald, N. Kunz, N.S. Merle, E.~E. West, P. Singh, A. Larochelle, F. Chinian, S. Mukherjee, B. Afzali, C. Kemper and N.~R. Zhang,
		\newblock Systematic single-cell pathway analysis to characterize early T cell activation,
		\newblock {\em Cell Reports}, 41(8):111697, 2022.
		
		\bibitem{smirnov}
		N. Smirnov, On the estimation of the discrepancy between empirical curves of distribution for two independent samples, {\it Bulletin de Universite de Moscow, Serie internationale (Mathematiques)}, Vol. 2, 3--14, 1939.
		
		
		\bibitem{tib96}
		R. Tibshirani
		\newblock Regression shrinkage and selection via the lasso,
		\newblock {\em Journal of the Royal Statistical Society B}, 58:267–-2889, 1996.
		
		\bibitem{ww}
		A.~Wald and J.~Wolfowitz,
		\newblock On a test whether two samples are from the same population, 
		\newblock {\em Annals of Mathematical Statistics}, 11(2):147--162, 06 1940.
		
		\bibitem{weiss}
		L. Weiss, Two-sample tests for multivariate distributions, {\it The Annals of Mathematical Statistics}, Vol. 31, 159--164, 1960.
		
		\bibitem{schi}
		M.F. Schilling,  Multivariate two-sample tests based on nearest neighbors, {\it Journal of the American
			Statistical Association}, Vol. 81, 799-806, 1986.
		
		\bibitem{portugalpaper}
		P. Cortez and A. Morais,  A Data Mining Approach to Predict Forest Fires using Meteorological Data, {\it J. Neves, M. F. Santos and J. Machado Eds., New Trends in Artificial Intelligence, Proceedings of the 13th EPIA 2007 - Portuguese Conference on Artificial Intelligence}, 512-523, 2007.
		
		\bibitem{slovin}
		S. Slovin, A. Carissimo, F. Panariello, A. Grimaldi, V. Bouch{\'e}, G. Gambardella, D. Cacchiarelli,
		\newblock Single-Cell RNA Sequencing Analysis: A Step-by-Step Overview.
		\newblock {{\it RNA Bioinformatics}, Springer US, New York, NY}.
		\newblock{343-365, 2021}.
		
		\bibitem{bbab34}
		K. Su, T. Yu and H. Wu, Accurate feature selection improves single-cell RNA-seq cell clustering,
		\newblock {\em Briefings in Bioinformatics}, ISSN. 1477--4054 (2), 2021.
		
		\bibitem{hicks}
		F.W. Townes, S.C. Hicks, M.J. Aryee, R.A. Irizarry, Feature selection and dimension reduction for single-cell RNA-Seq based on a multinomial model, {\it Genome Biology}, Vol. 20, 295, 2019. 
		
		\bibitem{Tmetab}
		J.A. Shyer, R.A. Flavell and W. Bailis, Metabolic signaling in T cells, {\it Cell Research}, Vol. 30, 649-659, 2020.
		
		\bibitem{ldha}
		N. Jones, J.G. Cronin, G. Dolton, S. Panetti, A.J. Schauenburg, S.A.E. Galloway, A.K. Sewell, D.K. Cole, C.A. Thornton, N.J. Francis, Metabolic Adaptation of Human CD4+ and CD8+ T-Cells to T-Cell Receptor-Mediated Stimulation, {\it Frontiers in Immunology}, Vol. 8, 1561, 2017.
		
		\bibitem{b35} 
		A. M. Griesinger, D. K. Birks, A. M. Donson, V. Amani, L.M. Hoffman, A. Waziri, M. Wang, M. H. Handler and N. K. Foreman, Characterization of distinct immunophenotypes across pediatric brain tumor types. \textit{Journal of Immunology}, 191(9), 4880-4888, 2013.
		
		\bibitem{b36} P. Langfelder, S. Horvath. WGCNA: an R package for weighted correlation network analysis. \textit{BMC Bioinformatics}. 2008;9:559. doi: 10.1186/1471-2105-9-559.
		
		\bibitem{b37} BR. Zeeberg, W. Feng, G. Wang, et al. GoMiner: a resource for biological interpretation of genomic and proteomic data[J]. \textit{Genome biology}, 2003, 4(4): R28.
	\end{thebibliography}
\end{document}